\documentclass[sn-basic,Numbered]{sn-jnl}
\usepackage{booktabs,longtable,array,multirow,float}
\usepackage[table]{xcolor}
\usepackage{graphicx,amsmath,amssymb,amsfonts,amsthm,mathrsfs}
\usepackage[title]{appendix}
\usepackage{textcomp,manyfoot,algorithm,algorithmicx,algpseudocode}
\usepackage{dsfont,geometry,natbib,hyperref,bm}
\usepackage{caption}
\usepackage{subcaption}
\usepackage{tikz}
\usetikzlibrary{positioning,arrows.meta}
\usepackage{placeins}
\usepackage{needspace}

\theoremstyle{thmstyleone}
\newtheorem{theorem}{Theorem}
\newtheorem{proposition}[theorem]{Proposition}

\theoremstyle{thmstyletwo}

\theoremstyle{thmstylethree}

\begin{document}

\title[Pricing Structural Longevity Risk]
      {A Solvency II-Based Framework for Pricing Structural Longevity Risk under an Affine Mortality Model}

\author*[1]{\fnm{Madiha} \sur{Bouhouche}}\email{m.bouhouche@insea.ac.ma}
\author[1,2]{\fnm{Khalil} \sur{Said}}\email{khalil.said@univ-pau.fr}
\author[1]{\fnm{Fadoua} \sur{Badaoui}}\email{fbadaoui@insea.ac.ma}
 \affil[1]{
 	\orgname{National Institute of Statistics and Applied Economics, Rabat}, \orgaddress{\country{Morocco}}}
\affil[2]{
	\orgname{Université de Pau et des Pays de l’Adour, E2S UPPA, TREE, Pau},
	\orgaddress{\country{France}}} 
\abstract{%
	This paper develops a Solvency II-based pricing framework for structural longevity risk in incomplete longevity markets. The valuation scheme relies on the Cost-of-Capital principle underlying the Risk Margin and combines it with a continuous-time affine mortality model. Mortality dynamics are specified in calendar time, with age-dependent mortality patterns linked to a latent longevity factor driving long-term mortality improvements. The affine structure preserves closed-form survival probabilities and provides a tractable basis for repeated Best Estimate, Solvency Capital Requirement and Risk Margin calculations. We derive the conditional distribution of one-year discounted Best Estimate variations, obtaining analytical expressions for their asymmetry and Value-at-Risk under the Gaussian affine specification. Within this setting, we introduce a stress-based assessment by applying adverse shocks to the deterministic parameters governing the longevity trend. The trend-based stress is compared with a mortality-level stress and with a Value-at-Risk-based capital assessment. The methodology links stochastic mortality modelling, prudential capital valuation and longevity-linked payoff assessment within a unified actuarial framework. A numerical application based on Belgian mortality data illustrates the implementation and shows that the mortality-level stress produces substantially larger scenario-based capital-cost measures than the trend-based stress. These findings highlight the sensitivity of longevity risk valuation to the specification of adverse mortality scenarios. The approach offers a transparent and operational basis for assessing and pricing structural longevity risk under a Solvency II Cost-of-Capital perspective.	
}




\maketitle

\section*{Introduction}\label{sec:intro}

The persistent improvement in life expectancy has become a central source of risk for life insurers and pension funds. Longevity risk is usually decomposed into two components. The idiosyncratic component arises from random fluctuations in individual lifetimes and can be largely diversified in sufficiently large portfolios. The systematic component is driven by aggregate mortality improvements, affects entire cohorts simultaneously, and cannot be eliminated through pooling. This second component is particularly relevant for annuity providers, since unexpected improvements in survival probabilities increase the duration and value of future benefit payments, as emphasized by Blake and Burrows (2001)\cite{BlakeBurrows2001} and Cairns et al. (2006)\cite{CairnsBlakeDowd2006}.\\

The development of longevity-linked securities has been motivated by the need to transfer this systematic component to financial or reinsurance markets. Instruments such as survivor bonds, S-forwards and longevity swaps are designed to generate cash flows linked to realized survival or mortality outcomes. When longevity improvements exceed initial expectations, the increase in annuity liabilities may therefore be partly offset by gains on such instruments. This line of financial innovation is closely related to the contributions of Coughlan et al. (2007)\cite{Coughlan2007} and Dowd et al. (2006)\cite{Dowd2006}.\\

The pricing of longevity risk has been addressed through several methodological perspectives. A first stream relies on financial valuation principles and, in particular, on risk-neutral or martingale-based arguments adapted to stochastic mortality settings. This approach has been developed, among others, by Dahl (2004)\cite{Dahl2004}. Since longevity exposures are only imperfectly hedgeable, related approaches for incomplete insurance markets introduce explicit risk loadings or alternative valuation principles. The instantaneous Sharpe-ratio framework proposed by Young (2008)\cite{Young2008} is a prominent example of this line of work. Other contributions incorporate risk premia through actuarial or distortion-based methods, as in Wang (2000)\cite{Wang2000} and Denuit et al. (2005)\cite{Denuit2005}.\\

A second stream adopts a regulatory valuation perspective. Under Solvency II, the Cost-of-Capital framework underlying the Risk Margin provides a natural actuarial basis for valuing non-hedgeable risks over the lifetime of insurance obligations. In this setting, the Solvency Capital Requirement is calibrated to a 99.5\% Value-at-Risk criterion over a one-year horizon, thereby linking adverse one-year liability variations to prudential capital requirements. This regulatory perspective is consistent with the Solvency II valuation framework and with the approach developed by Levantesi and Menzietti (2017)\cite{LevantesiMenzietti2017}. The numerical application specifies a flat risk-free discount rate as a simplifying valuation assumption.\\

Although Value-at-Risk provides a standard measure of short-term solvency risk, it does not explicitly distinguish between temporary mortality fluctuations and persistent changes in long-term mortality trends. This distinction is important for longevity risk, because annuity liabilities are particularly sensitive to structural improvements in survival probabilities. Stress-based methods therefore offer a complementary perspective by assessing the sensitivity of liabilities to targeted changes in the underlying mortality assumptions. Depending on the nature of the adverse scenario, stress-based valuation and VaR-based capital assessments may highlight different aspects of longevity risk, as discussed by Gylys and \v{S}iaulys (2019)\cite{GylysSiaulys2019}.\\

The purpose of this paper is to develop a Solvency II-consistent pricing framework for structural longevity risk. The proposed approach combines a Cost-of-Capital valuation principle with a continuous-time affine mortality model and focuses on longevity exposures whose value depends on deviations between stochastic survival dynamics and deterministic reference assumptions. The framework is designed to connect stochastic mortality modelling, Best Estimate valuation, one-year Solvency Capital Requirement computation and Risk Margin assessment in a unified actuarial setting.\\

The paper makes two contributions to the actuarial valuation of longevity-linked exposures. First, building on the calendar-year affine mortality model of Hainaut (2023) \cite{Hainaut2023}, we characterize the conditional distribution of one-year discounted Best Estimate variations for a longevity-linked payoff. This yields an explicit expression for the conditional Value-at-Risk and allows the projected capital requirements entering the Cost-of-Capital Risk Margin to be evaluated from expected future conditional SCRs. The contribution thus connects analytical mortality dynamics with the one-year revaluation risk of the liability and its capital-cost assessment.\\

Second, we develop a scenario-based assessment of structural longevity risk by applying prospective shocks to the deterministic reversion target of the latent longevity factor. In contrast to a direct mortality-level shock, this specification changes the assumptions governing future mortality dynamics while retaining the central mortality state at each valuation date. The resulting scenario-based capital-cost measures are compared with the analytical VaR-based assessment to illustrate the effects of distinct adverse mortality mechanisms.\\

The rest of the paper is organized in four main sections. Section~\ref{sec:affine_mortality} introduces the affine mortality framework, the closed-form survival probabilities, Gaussian calibration and predictive validation. Section~\ref{sec:solvency_pricing} develops the Solvency II-consistent pricing framework, including the longevity-linked payoff, the Best Estimate and the one-year SCR-based Risk Margin. Section~\ref{sec:structural_stress} presents the mortality-level and trend-based stress specifications, together with the associated scenario-based capital construction. Section~\ref{sec:empirical_application} reports the empirical application to Belgian mortality data, including calibration, out-of-sample validation, VaR-based capital assessment, stress-based pricing results and numerical stability checks.

\section{Affine mortality framework and survival probabilities}
\label{sec:affine_mortality}

The valuation of structural longevity risk requires a mortality model that is sufficiently rich to represent persistent changes in survival patterns, while remaining analytically tractable for repeated liability and capital calculations. This section introduces the continuous-time affine mortality framework used throughout the paper. The model combines an age-dependent mortality profile with a latent longevity factor governing aggregate mortality improvements. Its affine structure leads to closed-form survival probabilities, which provide the analytical basis for the Best Estimate, Solvency Capital Requirement and Risk Margin computations developed in the following sections.

\subsection{Continuous-time affine mortality model}
\label{subsec:continuous_affine_model}

The modelling of stochastic mortality dynamics has increasingly been shaped by the need to quantify longevity risk and to value mortality-linked liabilities and securities. For such applications, mortality models must capture realistic prospective survival patterns while maintaining a tractable link between mortality dynamics, survival probabilities and liability valuation. Affine diffusion processes are particularly well suited to this purpose, as they combine stochastic flexibility with analytical tractability in a way that is closely related to term-structure modelling in finance.\\

Early contributions established affine and mean-reverting processes as a natural framework for stochastic mortality modelling. Milevsky and Promislow (2001)\cite{MilevskyPromislow2001} introduced mean-reverting affine processes for mortality modelling, while Luciano and Vigna (2005)\cite{LucianoVigna2005} developed a jump-diffusion affine framework and highlighted the limitations of constant mean-reversion structures in mortality applications. Biffis (2005)\cite{Biffis2005} further exploited the tractability of affine processes for the valuation of life insurance liabilities and longevity-linked contracts, and Cairns et al. (2006)\cite{CairnsBlakeDowd2006} investigated the pricing of mortality-linked securities through adaptations of arbitrage-free interest-rate pricing methodologies. Schrager (2006)\cite{Schrager2006} proposed a multivariate affine framework calibrated to Dutch mortality data, while Luciano and Vigna (2008)\cite{LucianoVigna2008} examined the empirical relevance of affine specifications across generations of the UK population.\\

Subsequent developments extended stochastic mortality models to richer structures incorporating cohort effects, mortality surfaces, dependence across populations, stochastic shocks and solvency applications. Hainaut and Devolder (2008)\cite{HainautDevolder2008} proposed a cohort mortality model based on a mean-reverting L\'evy process, and Jevti\'c et al. (2013)\cite{JevticLucianoVigna2013} introduced a continuous-time factor approach designed to reproduce the main characteristics of mortality surfaces. Chen et al. (2015)\cite{Chen2015} developed a multi-population mortality framework based on a factor copula approach. Zeddouk and Devolder (2020)\cite{ZeddoukDevolder2020} proposed a dependent multi-cohort framework for longevity modelling and pricing applications, while Xu et al. (2020)\cite{XuSherrisZiveyi2020} developed a continuous-time multi-cohort mortality framework incorporating both common and cohort-specific stochastic factors. More recent contributions have also emphasized additional sources of uncertainty in mortality projections. Bravo and Nunes (2021)\cite{Bravo2021} developed a Fourier-transform approach for pricing longevity derivatives under affine jump-diffusion mortality models, Robben and Antonio (2024)\cite{RobbenAntonio2023} incorporated catastrophe risk through regime-switching mortality shocks in a stochastic multi-population framework, and Zhu and Bauer (2022)\cite{ZhuBauer2022} proposed a forward-looking stochastic framework for modelling uncertainty in long-term mortality projections.\\

Within this literature, affine and mean-reverting mortality models remain especially attractive for pricing and regulatory valuation because they link stochastic mortality dynamics to explicit or semi-explicit survival and valuation quantities. In this paper, we rely on the continuous-time calendar-year mortality model introduced by Hainaut (2023)\cite{Hainaut2023}. Unlike classical cohort-based affine specifications, which typically associate one mortality process with each generation, this model indexes mortality intensities jointly by age and calendar time. This representation preserves analytical tractability, facilitates calibration over relatively short observation windows, and naturally induces dependence across cohorts through a common longevity factor.\\

In this setting, mortality is described as a mean-reverting random field driven by a latent factor governing aggregate longevity improvements. The latent factor explicitly represents the long-term mortality trend, which makes it possible to apply longevity stresses directly to the deterministic parameters governing future mortality improvements rather than to the mortality surface as a whole. This link between the latent trend and future survival projections is central to the stress-based pricing methodology developed in this paper.\\

Following Hainaut (2023)\cite{Hainaut2023}, let $\mu_{t,x}$ denote the instantaneous mortality intensity at calendar time $t$ for an individual aged $x$. The mortality dynamics are given by
\begin{equation}\label{eq:mu_dyn}
	d\mu_{t,x}
	=
	\kappa\big(\theta_t\,\mu(x)-\mu_{t,x}\big)\,dt
	+\Sigma(x)^\top dW_t^\mu,
\end{equation}
where $\kappa>0$ is the speed of mean reversion, $\mu(x)$ is a deterministic baseline mortality profile, $\Sigma(x)\in\mathbb{R}^d$ is an age-dependent volatility structure, and $W_t^\mu$ is a $d$-dimensional Brownian motion. The mortality intensity therefore fluctuates around a stochastic reversion level determined by the baseline age profile and by the longevity factor $\theta_t$, while the volatility structure allows mortality uncertainty to vary with age.\\

The latent longevity factor $(\theta_t)_{t\ge 0}$ follows
\begin{equation}\label{eq:theta_dyn}
	d\theta_t
	=
	\alpha\big(\beta(t)-\theta_t\big)\,dt
	+\nu\,dW_t^\theta,
\end{equation}
where $\alpha>0$ is the speed of reversion, $\nu>0$ is the volatility parameter, and $W_t^\theta$ is a one-dimensional Brownian motion. The deterministic reversion target is specified as
\begin{equation}\label{eq:beta_def}
	\beta(t)=\theta_0e^{-\gamma t}+\bar\theta\big(1-e^{-\gamma t}\big),
	\qquad
	\theta_0>\bar\theta>0,\quad \gamma>0.
\end{equation}
This specification represents a progressive longevity improvement over calendar time, with convergence towards a long-run level. Compared with affine mortality models based on constant long-term means, the time-dependent target $\beta(t)$ provides additional flexibility for representing persistent changes in mortality trends.\\

Let $\mathcal{F}_t=\mathcal{G}_t\vee\mathcal{H}_t$ denote the filtration of the model, where $\mathcal{G}_t$ is generated by the mortality process and $\mathcal{H}_t$ by the latent factor $\theta_t$. Following the original specification, the deterministic baseline mortality curve is assumed to follow a Gompertz--Makeham law,
\begin{equation*}
	\mu(x)=a+bc^x.
\end{equation*}
Although dependence between $W^\mu$ and $W^\theta$ could be considered theoretically, the latent nature of $\theta_t$ makes such dependence difficult to identify empirically. For calibration purposes, the Brownian motions $W^\mu$ and $W^\theta$ are therefore assumed to be independent.

\subsection{Closed-form survival probabilities}
\label{subsec:survival_probabilities}

Let ${}_{\xi}p_{t,x}$ denote the survival probability, evaluated at calendar time $t$, that an individual aged $x$ survives over the horizon $\xi=s-t$. Under the intensity-based specification, this probability is given by
\begin{equation*}
	{}_{\xi}p_{t,x}
	=
	\mathbb{E}\left[
	\exp\left(
	-\int_0^{\xi}\mu_{t+u,x+u}\,du
	\right)
	\,\middle|\,\mathcal{F}_t
	\right].
\end{equation*}

Following Hainaut (2023)\cite{Hainaut2023}, the affine structure of the model yields the closed-form representation
\begin{equation}\label{eq:survival_cf}
	{}_{\xi}p_{t,x}
	=
	\exp\left(
	-{}_{\xi}m_{t,x}
	+\frac{1}{2}{}_{\xi}v_{t,x}
	\right),
	\qquad \xi=s-t,
\end{equation}
where ${}_{\xi}m_{t,x}$ and ${}_{\xi}v_{t,x}$ denote, respectively, the conditional mean and variance components of the integrated mortality intensity over the projection horizon. In the original framework, the mortality dynamics may include both diffusion and jump components. The present paper focuses on the purely diffusive specification, which preserves the analytical tractability required for repeated valuation while providing a parsimonious setting for calibration and stress testing.\\

Under this specification, the drift and variance components are given by
\begin{equation}\label{eq:mv_general}
    \begin{aligned}
        {}_\xi m_{t,x}
        &=
        \int_0^\xi e^{-\kappa u}\mu_{t,x+u}\,du
        +
        \frac{\kappa\theta_t}{\kappa-\alpha}
        \int_0^\xi \mu(x+u)
        \left(e^{-\alpha u}-e^{-\kappa u}\right)\,du \\
        &\quad
        +
        \frac{\kappa\alpha(\theta_0-\bar{\theta})e^{-\gamma t}}
        {\kappa-\alpha}
        \int_0^\xi \mu(x+u)
        \left(
        \frac{e^{-\gamma u}-e^{-\alpha u}}{\alpha-\gamma}
        -
        \frac{e^{-\gamma u}-e^{-\kappa u}}{\kappa-\gamma}
        \right)\,du
        \\
        &\quad
        +
        \frac{\kappa\alpha\bar{\theta}}{\kappa-\alpha}
        \int_0^\xi \mu(x+u)
        \left(
        \frac{1-e^{-\alpha u}}{\alpha}
        -
        \frac{1-e^{-\kappa u}}{\kappa}
        \right)\,du,
        \\[1mm]
        {}_\xi v_{t,x}
        &=
        \left(\frac{\kappa\nu}{\kappa-\alpha}\right)^2
        \int_0^\xi
        \left(
        \int_z^\xi \mu(x+u)
        \left(e^{-\alpha(u-z)}-e^{-\kappa(u-z)}\right)\,du
        \right)^2 dz \\
        &\quad
        +
        \int_0^\xi
        \left(
        \int_z^\xi \Sigma(x+u)e^{-\kappa(u-z)}\,du
        \right)^\top
        \left(
        \int_z^\xi \Sigma(x+u)e^{-\kappa(u-z)}\,du
        \right)\,dz.
    \end{aligned}
\end{equation}

Under the Gompertz--Makeham specification
$\mu(x)=a+bc^x$, define the auxiliary integral
\[
J_\lambda(x,\xi)
=
\int_0^\xi \mu(x+u)e^{-\lambda u}\,du.
\]
Direct integration gives
\[
J_\lambda(x,\xi)
=
a\frac{1-e^{-\lambda\xi}}{\lambda}
+
bc^x
\frac{e^{(\ln c-\lambda)\xi}-1}
{\ln c-\lambda},
\]
with the corresponding continuous extensions
when a denominator vanishes.

Using this notation, the conditional mean
in~\eqref{eq:mv_general} admits the following
compact analytical representation:
\begin{equation}\label{eq:m_explicit}
    \begin{aligned}
        {}_\xi m_{t,x}
        ={}&
        \int_0^\xi e^{-\kappa u}
        \mu_{t,x+u}\,du
        \\[2pt]
        &+
        \frac{\kappa\theta_t}{\kappa-\alpha}
        \left(J_\alpha-J_\kappa\right)
        \\[2pt]
        &+
        \frac{\kappa\alpha(\theta_0-\bar{\theta})
        e^{-\gamma t}}{\kappa-\alpha}
        \left[
        \frac{J_\gamma-J_\alpha}{\alpha-\gamma}
        -
        \frac{J_\gamma-J_\kappa}{\kappa-\gamma}
        \right]
        \\[2pt]
        &+
        \frac{\kappa\alpha\bar{\theta}}{\kappa-\alpha}
        \left[
        \frac{J_0-J_\alpha}{\alpha}
        -
        \frac{J_0-J_\kappa}{\kappa}
        \right],
    \end{aligned}
\end{equation}
where $J_\lambda=J_\lambda(x,\xi)$.

For the conditional variance, we use the auxiliary function
introduced in Hainaut (2023) \cite{Hainaut2023},
\[
\begin{aligned}
h(\xi;\beta_1,\beta_2)
={}&
\xi
+
\frac{1-e^{-(\beta_1+\beta_2)\xi}}
{\beta_1+\beta_2}
\\
&-
\frac{1-e^{-\beta_2\xi}}{\beta_2}
-
\frac{1-e^{-\beta_1\xi}}{\beta_1},
\end{aligned}
\]
with the convention
$h(\xi;\beta)=h(\xi;\beta,\beta)$.

Under the Gompertz--Makeham specification, the
age-dependent mortality profile introduces an additional
exponential weighting in the variance calculation.
To account for this effect, we extend the auxiliary
function to
\[
h_\rho(\xi;\beta_1,\beta_2)
=
\int_0^\xi
e^{\rho z}
\left(1-e^{-\beta_1(\xi-z)}\right)
\left(1-e^{-\beta_2(\xi-z)}\right)
\,dz.
\]

This generalized function admits the closed-form
representation
\[
\begin{aligned}
h_\rho(\xi;\beta_1,\beta_2)
={}&
\frac{e^{\rho\xi}-1}{\rho}
\\
&-
\frac{e^{\rho\xi}-e^{-\beta_1\xi}}
{\rho+\beta_1}
-
\frac{e^{\rho\xi}-e^{-\beta_2\xi}}
{\rho+\beta_2}
\\
&+
\frac{e^{\rho\xi}-e^{-(\beta_1+\beta_2)\xi}}
{\rho+\beta_1+\beta_2},
\end{aligned}
\]
where vanishing denominators are handled by their
continuous limits. In particular,
\[
h_0(\xi;\beta_1,\beta_2)
=
h(\xi;\beta_1,\beta_2).
\]

Writing $\lambda=\ln c$, the conditional variance
can then be expressed as

\begin{align}\label{eq:v_explicit}
{}_\xi v_{t,x}
={}&
\left(\frac{\kappa\nu}{\kappa-\alpha}\right)^2
\Bigg[
a^2\left(
\frac{h_0(\xi;\alpha,\alpha)}{\alpha^2}
+\frac{h_0(\xi;\kappa,\kappa)}{\kappa^2}
-\frac{2h_0(\xi;\alpha,\kappa)}{\alpha\kappa}
\right)
\notag\\
&\quad
+(bc^x)^2\Bigg(
\frac{h_{2\lambda}(\xi;\alpha-\lambda,\alpha-\lambda)}
{(\lambda-\alpha)^2}
+\frac{h_{2\lambda}(\xi;\kappa-\lambda,\kappa-\lambda)}
{(\lambda-\kappa)^2}
\notag\\
&\qquad
-\frac{2h_{2\lambda}(\xi;\alpha-\lambda,\kappa-\lambda)}
{(\lambda-\alpha)(\lambda-\kappa)}
\Bigg)
\notag\\
&\quad
+2abc^x\Bigg(
-\frac{h_\lambda(\xi;\alpha,\alpha-\lambda)}
{\alpha(\lambda-\alpha)}
+\frac{h_\lambda(\xi;\alpha,\kappa-\lambda)}
{\alpha(\lambda-\kappa)}
\notag\\
&\qquad
+\frac{h_\lambda(\xi;\kappa,\alpha-\lambda)}
{\kappa(\lambda-\alpha)}
-\frac{h_\lambda(\xi;\kappa,\kappa-\lambda)}
{\kappa(\lambda-\kappa)}
\Bigg)
\Bigg]
\notag\\
&\quad
+\int_0^\xi
\left(
\int_z^\xi \Sigma(x+u)e^{-\kappa(u-z)}\,du
\right)^\top
\left(
\int_z^\xi \Sigma(x+u)e^{-\kappa(u-z)}\,du
\right)\,dz.
\end{align}

The generalized auxiliary function therefore provides
an exact analytical evaluation of the latent-factor
variance contribution under the Gompertz--Makeham
mortality profile. The terms involving $a^2$ retain
the original auxiliary function, whereas the mixed
terms involving $abc^x$ and the quadratic terms
involving $(bc^x)^2$ require exponential weights
$\rho=\lambda$ and $\rho=2\lambda$, respectively.
The age-dependent volatility contribution remains
unchanged.

The final term in \eqref{eq:v_explicit} corresponds to the contribution of the age-dependent volatility structure. In a multidimensional setting, this contribution does not generally reduce to a simple explicit expression. In the empirical application, the principal component analysis of historical mortality variations supports the use of a one-factor specification for mortality volatility. We therefore consider the exponential structure
\[
\Sigma(x)=\sigma_0 e^{\sigma_1 x},
\]
which preserves the closed-form nature of the survival probability while allowing mortality volatility to increase with age.\\

Define
\[
I(\xi,x)
=
\int_0^\xi
\left(
\int_z^\xi \Sigma(x+u)e^{-\kappa(u-z)}\,du
\right)^2 dz.
\]
For $\Sigma(x)=\sigma_0e^{\sigma_1x}$, the inner integral is
\[
\int_z^\xi
\sigma_0 e^{\sigma_1(x+u)}e^{-\kappa(u-z)}\,du
=
\sigma_0 e^{\sigma_1 x}e^{\sigma_1 z}
\frac{e^{(\sigma_1-\kappa)(\xi-z)}-1}{\sigma_1-\kappa}.
\]
Hence,
\[
I(\xi,x)
=
\frac{\sigma_0^2e^{2\sigma_1 x}}{(\sigma_1-\kappa)^2}
\int_0^\xi
e^{2\sigma_1 z}
\left(
e^{(\sigma_1-\kappa)(\xi-z)}-1
\right)^2 dz.
\]
After simplification, the volatility contribution is given by
\begin{equation}\label{eq:I_sigma_explicit}
	I(\xi,x)
	=
	\frac{\sigma_0^2 e^{2\sigma_1 x}}{(\sigma_1-\kappa)^2}
	\left[
	\frac{e^{2\sigma_1\xi}-e^{2(\sigma_1-\kappa)\xi}}{2\kappa}
	-
	2\frac{e^{2\sigma_1\xi}-e^{(\sigma_1-\kappa)\xi}}{\sigma_1+\kappa}
	+
	\frac{e^{2\sigma_1\xi}-1}{2\sigma_1}
	\right].
\end{equation}
Equations \eqref{eq:survival_cf}--\eqref{eq:I_sigma_explicit} provide the analytical basis for the repeated survival, Best Estimate and capital calculations used in the Solvency II-consistent pricing framework developed below.

\subsection{Gaussian calibration and predictive validation}
\label{subsec:observation_validation}

The estimation procedure relies on the one-year transition distribution implied by the affine mortality framework. Using the analytical solutions derived by Hainaut (2023)\cite{Hainaut2023}, the joint dynamics of the mortality intensity and the latent longevity factor admit an explicit Gaussian representation over one-year horizons. Since the longevity factor $\theta_t$ is not directly observable, it is estimated through a Gaussian filtering procedure based on the conditional distribution of mortality intensities. The model parameters are then obtained by maximizing the associated Gaussian log-likelihood.\\

For numerical calibration, the mortality-diffusion covariance matrix is approximated by its diagonal, while the dependence induced by the latent longevity factor is retained. This working-covariance approximation preserves the age-specific marginal variances implied by the estimated volatility profile and facilitates stable Gaussian likelihood estimation. The subsequent continuous-time survival valuation relies on the common-factor mortality specification introduced in Subsection~\ref{subsec:survival_probabilities}.\\

To assess predictive performance, out-of-sample validation is conducted
directly in the mortality-intensity space, consistently with the multivariate
Gaussian specification used for calibration and forecasting. Predictive validation is performed directly on mortality intensities, without introducing a separate observation model for annual death counts. For each
validation year $t$, the point forecast is defined as the empirical mean of
the simulated mortality intensities:
\begin{equation*}
    \widehat{\mu}_{t,x}
    =
    \frac{1}{N}
    \sum_{s=1}^{N}
    \mu_{t,x}^{(s)},
\end{equation*}
where $N$ denotes the number of simulated trajectories. Pointwise 95\%
predictive intervals are obtained from the empirical 2.5\% and 97.5\%
quantiles of the simulated mortality intensities.

Since prediction errors at highly exposed ages have a larger actuarial impact,
predictive accuracy is assessed using exposure-weighted error measures. For
each validation year $t$, the exposure weight associated with age $x$ is
defined as
\begin{equation*}
    w_{t,x}
    =
    \frac{E_{t,x}}
    {\sum_{x\in\mathcal{A}}E_{t,x}},
    \qquad x\in\mathcal{A},
\end{equation*}
where $\mathcal{A}$ denotes the set of ages. The weighted mean absolute error
and weighted root mean squared error are given by
\begin{equation*}
\begin{aligned}
    \operatorname{MAE}_{w,t}
    &=
    \sum_{x\in\mathcal{A}}
    w_{t,x}
    \left|
    \widehat{\mu}_{t,x}
    -
    \mu^{\mathrm{obs}}_{t,x}
    \right|,\\
    \operatorname{RMSE}_{w,t}
    &=
    \left[
    \sum_{x\in\mathcal{A}}
    w_{t,x}
    \left(
    \widehat{\mu}_{t,x}
    -
    \mu^{\mathrm{obs}}_{t,x}
    \right)^2
    \right]^{1/2}.
\end{aligned}
\end{equation*}

The empirical coverage of the pointwise 95\% predictive intervals is also
reported:
\begin{equation*}
    \operatorname{Coverage}_{0.95,t}
    =
    \frac{1}{|\mathcal{A}|}
    \sum_{x\in\mathcal{A}}
    \mathbf{1}
    \left\{
    L_{t,x}^{0.025}
    \leq
    \mu^{\mathrm{obs}}_{t,x}
    \leq
    U_{t,x}^{0.975}
    \right\},
\end{equation*}
where $L_{t,x}^{0.025}$ and $U_{t,x}^{0.975}$ denote the lower and upper
predictive quantiles.

Model-consistent predictive residuals are defined by
\begin{equation*}
    r^{G}_{t,x}
    =
    \frac{
    \mu^{\mathrm{obs}}_{t,x}
    -
    \widehat{\mu}_{t,x}
    }{
    s^{\mathrm{pred}}_{t,x}
    },
\end{equation*}
where $s^{\mathrm{pred}}_{t,x}$ is the empirical standard deviation of the
simulated mortality intensities. Under a well-calibrated marginal Gaussian predictive distribution,
approximately 95\% of standardized residuals are expected to lie
within the reference bounds of $\pm 1.96$, when considered marginally. These thresholds are pointwise and are not interpreted as simultaneous
confidence limits across all ages.

These validation criteria are used to assess whether the calibrated affine
mortality model provides a sufficiently stable empirical basis for the
valuation, capital and stress-testing exercises developed in the sequel.
The empirical results are reported in
Subsection~\ref{subsec:out_of_sample_validation}.

\section{Solvency II-consistent pricing framework}
\label{sec:solvency_pricing}

The affine mortality framework introduced in Section~\ref{sec:affine_mortality} provides the survival quantities required for the valuation of longevity-linked exposures. This section develops the corresponding Solvency II-consistent pricing framework. The approach is based on the Cost-of-Capital principle underlying the Risk Margin and is designed to value non-hedgeable structural longevity risk in an incomplete market setting. We first discuss the regulatory valuation rationale, then introduce the longevity-linked payoff and its Best Estimate, before deriving the one-year SCR and Risk Margin representation used in the subsequent stress-based analysis.\\

\subsection{Market incompleteness and Cost-of-Capital valuation}
\label{subsec:market_incompleteness_coc}

Longevity-linked liabilities are exposed to systematic mortality improvements that cannot be fully diversified through portfolio pooling or perfectly replicated by traded financial assets. Although longevity risk transfer markets have developed through reinsurance transactions and longevity-linked instruments, they remain relatively illiquid and concentrated among a limited number of specialized participants, as discussed by Blake and Cairns (2020)\cite{BlakeCairns2020} and B{\"o}rger et al. (2023)\cite{BorgerFreimannRuss2023}. This market incompleteness is central for valuation, since it prevents the identification of a unique arbitrage-free price for longevity-linked exposures.\\

In such a setting, a regulatory pricing approach based on capital costs provides a natural valuation principle. Under Solvency II, technical provisions are valued as the sum of a Best Estimate and a Risk Margin. The Best Estimate represents the probability-weighted present value of future cash flows, while the Risk Margin reflects the cost of holding capital against non-hedgeable risks over the lifetime of the liabilities. The corresponding decomposition is given by
\begin{equation*}
	\text{Liability Value}
	=
	\text{Best Estimate}
	+
	\text{Risk Margin}.
\end{equation*}

The Risk Margin is particularly relevant for longevity-linked exposures because systematic longevity risk cannot be eliminated by increasing portfolio size and is only imperfectly transferable through market instruments. Within the Cost-of-Capital framework, it measures the present value of future capital costs associated with bearing this non-hedgeable risk. This interpretation is consistent with the Solvency II valuation framework and with the approach of Levantesi and Menzietti (2017)\cite{LevantesiMenzietti2017}, who relate the Risk Margin to the maximum amount an insurer may be willing to pay to transfer longevity risk.\\

In the present paper, this Cost-of-Capital interpretation is used to construct a prudential pricing framework for structural longevity risk. The resulting price is not obtained through perfect replication, but through the capital cost generated by the stochastic evolution of longevity-linked liabilities. This provides a coherent actuarial basis for valuing longevity-linked payoffs in an incomplete market environment and prepares the Risk Margin methodology developed below.

\subsection{Longevity-linked payoff and Best Estimate}
\label{subsec:payoff_best_estimate}

Standard S-forward contracts are designed to transfer systematic longevity risk by exchanging a realized survival outcome against a deterministic survival benchmark fixed at inception. They belong to the broader class of longevity-linked securities developed for the transfer of mortality and longevity risk, including survivor bonds, q-forwards, S-forwards and longevity swaps \cite{BlakeBurrows2001,Coughlan2007,Dowd2006}. By contrast, S-exchange structures are mainly designed to address basis risk arising from discrepancies between the mortality dynamics of insured portfolios and those of reference populations \cite{ZeddoukDevolder2022}.\\

The instrument considered in this paper is closer in spirit to an S-forward. It is written on the deviation between stochastic survival dynamics and a deterministic regulatory reference. The payoff therefore captures the discrepancy between survival probabilities generated by the stochastic mortality model and those implied by a fixed technical or regulatory benchmark, rather than cross-population basis risk.\\

\begin{figure}[H]
	\centering
	\begin{tikzpicture}[>=Stealth, node distance=2cm]
		
		\node (stoch) [draw, rounded corners, align=center, minimum width=3.6cm, minimum height=1.2cm] 
		{Stochastic mortality\\ dynamics};
		
		\node (contract) [draw, rounded corners, right=2cm of stoch, align=center, minimum width=3.2cm, minimum height=1.2cm]
		{S-forward\\ payoff};
		
		\node (ref) [draw, rounded corners, right=2cm of contract, align=center, minimum width=3.6cm, minimum height=1.2cm]
		{Deterministic\\ reference table};
		
		\draw[->, thick] (stoch) -- node[above]{\(\widetilde I(x,T)\)} (contract);
		
		\draw[->, thick] (ref) -- node[above]{\({}_{T-t_0}\bar p_x\)} (contract);
		
		\node (payoff) [below=1cm of contract, align=center]
		{\(\displaystyle f(T)=\widetilde I(x,T)-{}_{T-t_0}\bar p_x\)};
		
	\end{tikzpicture}
	\caption{Payoff structure of the S-forward contract.}
	\label{fig:sforward_payoff}
\end{figure}
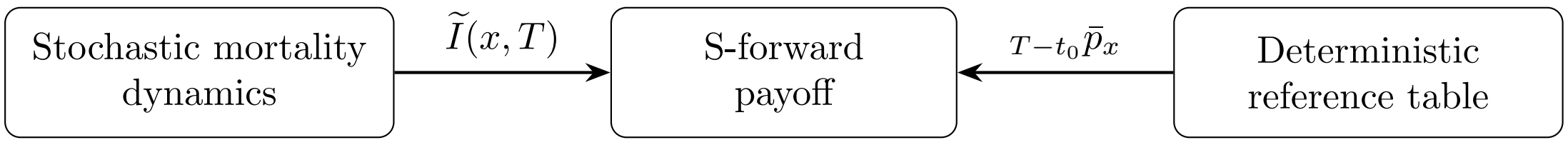

We consider a cohort of policyholders aged $x$ at the initial date $t_0$ and a maturity date $T>t_0$. The payoff of the longevity-linked contract is defined as
\begin{equation}\label{eq:sforward_payoff}
	f(T)
	=
	\widetilde I(x,T)
	-
	{}_{T-t_0}\bar p_x,
\end{equation}
where $\widetilde I(x,T)$ denotes the stochastic survival index generated by the mortality model, and ${}_{T-t_0}\bar p_x$ denotes the deterministic reference survival probability fixed at inception. The payoff structure is summarized in Figure~\ref{fig:sforward_payoff}.

The payoff is used here as a tractable survival-index representation of the discrepancy between stochastic mortality projections and the deterministic benchmark adopted in the regulatory valuation framework. It therefore captures the pricing impact of deviations in survival expectations while remaining analytically compatible with the Best Estimate and Risk Margin calculations developed below.\\

For notational simplicity, the initial date is set to $t_0=0$ in the sequel. The same formulas apply after a deterministic time shift. The stochastic survival index at maturity is given by
\begin{equation*}
	\widetilde I(x,T)
	=
	\exp\left(
	-\int_{0}^{T}
	\mu_{u,x+u}\,du
	\right).
\end{equation*}

At an intermediate date $i<T$, this index can be decomposed into the realized survival component over $[0,i]$ and the remaining stochastic survival component over $[i,T]$:
\begin{equation*}
	\widetilde I(x,T)
	=
	\widetilde I(x,i)
	\exp\left(
	-\int_{0}^{T-i}
	\mu_{i+u,x+i+u}\,du
	\right),
\end{equation*}
where
\begin{equation*}
	\widetilde I(x,i)
	=
	\exp\left(
	-\int_{0}^{i}
	\mu_{u,x+u}\,du
	\right)
\end{equation*}
is known at date $i$. Taking the conditional expectation with respect to $\mathcal F_i$ gives
\begin{equation*}
	\mathbb E\left[
	\widetilde I(x,T)
	\mid \mathcal F_i
	\right]
	=
	\widetilde I(x,i)\,
	{}_{T-i}p_{i,x+i}.
\end{equation*}

Using the closed-form survival probability derived in Subsection~\ref{subsec:survival_probabilities}, the conditional survival probability is
\begin{equation*}
	{}_{T-i}p_{i,x+i}
	=
	\exp\left(
	-{}_{T-i}m_{i,x+i}
	+
	\frac{1}{2}\,{}_{T-i}v_{i,x+i}
	\right).
\end{equation*}

It follows that the Best Estimate of the longevity-linked payoff at date $i$ is
\begin{equation}\label{eq:best_estimate_i}
	BE_i
	=
	P(i,T)
	\left(
	\widetilde I(x,i)\,
	{}_{T-i}p_{i,x+i}
	-
	{}_{T-t_0}\bar p_x
	\right),
\end{equation}
where $P(i,T)$ denotes the discount factor from maturity $T$ to date $i$. The sequence $(BE_i)_{i=0,\ldots,T}$ provides the valuation basis for the one-year liability variations and the SCR-based Risk Margin developed in Subsection~\ref{subsec:scr_risk_margin}.

\subsection{One-year SCR and Risk Margin valuation}
\label{subsec:scr_risk_margin}

The Risk Margin is computed from the sequence of one-year Solvency Capital Requirements generated by the stochastic revaluation of the longevity-linked payoff. On an annual projection grid, the initial Risk Margin is defined by
\begin{equation}\label{eq:risk_margin_def}
	RM_0
	=
	c_{\mathrm{CoC}}
	\sum_{i=0}^{T-1}
	SCR_{i\mid 0}\,
	P(0,i+1),
\end{equation}
where $c_{\mathrm{CoC}}$ denotes the Cost-of-Capital rate, $P(0,i+1)$ is the discount factor from date $i+1$ to date $0$, and $SCR_{i\mid 0}$ is the projected one-year capital requirement at date $i$, evaluated from the initial date. In the sequel, $SCR_i$ denotes the corresponding conditional capital requirement at projection date $i$.\\

\noindent In the numerical application, the primary projected capital requirement is $SCR_{i\mid0}=\mathbb E_0[SCR_i]$, where $SCR_i$ is the conditional VaR in \eqref{eq:scr_i_def}. The conditional SCR is characterized in Proposition~\ref{prop:conditional_be_var}, and its analytical projection to date $0$ is given in \eqref{eq:projected_scr_analytic}. We compare the resulting Risk Margin with a Monte Carlo estimate of the same expected conditional SCRs and, separately, with an ex-ante quantile of the unconditional one-year variations.

The one-year liability variation over the interval $[i,i+1]$ is defined as
\begin{equation}\label{eq:delta_i_def}
	\Delta_i
	=
	P(i,i+1)BE_{i+1}
	-
	BE_i.
\end{equation}

This quantity measures the change in the discounted Best Estimate induced by the updating of mortality information over one year. The associated Solvency Capital Requirement is given by
\begin{equation}\label{eq:scr_i_def}
	SCR_i
	=
	\operatorname{VaR}_{99.5\%}
	\left(
	\Delta_i
	\mid
	\mathcal F_i
	\right).
\end{equation}
Thus, the one-year capital requirement reflects the adverse revaluation of the longevity-linked payoff as survival expectations are revised under the stochastic mortality dynamics.\\

\begin{proposition}[One-year decomposition of the Best Estimate variation]
	\label{prop:delta_i_decomposition}
	Assume that discount factors are deterministic and that the Best Estimate at date $i$ is given by \eqref{eq:best_estimate_i}. For any projection date $i<T$, the one-year variation
	\[
	\Delta_i=P(i,i+1)BE_{i+1}-BE_i
	\]
	admits the decomposition
	\begin{equation}\label{eq:delta_i_decomposition}
		\Delta_i
		=
		P(i,T)A_i,
	\end{equation}
	where
	\begin{equation}\label{eq:A_i_def}
		A_i
		=
		\widetilde I(x,i)
		\left[
		\underbrace{
			\exp\left(
			-\int_{0}^{1}\mu_{i+u,x+i+u}\,du
			\right)
			{}_{T-(i+1)}p_{i+1,x+i+1}
		}_{C_i}
		-
		\underbrace{
			{}_{T-i}p_{i,x+i}
		}_{D_i}
		\right].
	\end{equation}
\end{proposition}

The term $C_i$ combines the realized survival over the interval $[i,i+1]$ with the updated conditional projection over the remaining horizon $(i+1,T)$. It therefore represents the revised valuation of future survival at date $i+1$. The term $D_i$ corresponds to the conditional projection available at date $i$. Consequently, the difference $C_i-D_i$ measures the revision of future survival expectations induced by the stochastic mortality dynamics over the one-year horizon.\\

Since the deterministic benchmark is fixed at inception, its contribution cancels out in the computation of $\Delta_i$. The one-year liability variation is therefore entirely driven by the stochastic updating of future survival expectations. Unlike standard insurance liabilities with recurrent cash flows, the S-forward payoff considered here is settled at maturity. Hence, no monotonicity property is expected in general for the sequence $(BE_i)_{i\geq 0}$.\\

Under the affine Gaussian specification, the conditional Value-at-Risk admits the analytical characterization established in Proposition~\ref{prop:conditional_be_var}. The empirical 99.5\% quantile of the pooled simulated one-year variations is retained only as a separate ex-ante, unconditional Monte Carlo measure.

\begin{proof}
	By definition,
	\[
	\Delta_i
	=
	P(i,i+1)BE_{i+1}-BE_i.
	\]
	Using the expression of the Best Estimate in \eqref{eq:best_estimate_i}, together with the deterministic discounting identity
	\[
	P(i,i+1)P(i+1,T)=P(i,T),
	\]
	we obtain
	\[
	\Delta_i
	=
	P(i,T)
	\left[
	\widetilde I(x,i+1){}_{T-(i+1)}p_{i+1,x+i+1}
	-
	\widetilde I(x,i){}_{T-i}p_{i,x+i}
	+
	{}_{T-t_0}\bar p_x
	-
	{}_{T-t_0}\bar p_x
	\right].
	\]
	Moreover,
	\[
	\widetilde I(x,i+1)
	=
	\widetilde I(x,i)
	\exp\left(
	-\int_{0}^{1}\mu_{i+u,x+i+u}\,du
	\right).
	\]
	Substituting this identity into the previous expression yields \eqref{eq:delta_i_decomposition} with $A_i$ defined by \eqref{eq:A_i_def}.
\end{proof}

\medskip


\medskip

The decomposition established in
Proposition~\ref{prop:delta_i_decomposition}
allows us to characterize the conditional distribution
of the one-year Best Estimate variation under the
affine Gaussian mortality model.

For each $i<T$, define the remaining integrated mortality
and the mortality accumulated over the following year by
\begin{equation*}
Y_i=\int_i^T \mu_{u,x+u}\,du,
\qquad
H_i=\int_i^{i+1}\mu_{u,x+u}\,du.
\end{equation*}

Let
\begin{equation*}
M_i=\mathbb E[Y_i\mid\mathcal F_i],
\qquad
V_i=\operatorname{Var}(Y_i\mid\mathcal F_i)
\end{equation*}
denote its conditional mean and variance.

Using the survival quantities appearing in
Proposition~\ref{prop:delta_i_decomposition},
we write
\begin{equation*}
\begin{aligned}
C_i
&=
e^{-H_i}\,
{}_{T-(i+1)}p_{i+1,x+i+1},\\
D_i
&=
{}_{T-i}p_{i,x+i}.
\end{aligned}
\end{equation*}

We further define
\begin{equation*}
\mathcal R_i
=
\log\left(\frac{C_i}{D_i}\right),
\qquad
B_i
=
P(i,T)\widetilde I(x,i)D_i>0.
\end{equation*}

The following proposition characterizes the conditional
distribution of the discounted Best Estimate variation.

\begin{proposition}
[Conditional distribution and Value-at-Risk of the
Best Estimate variation]
\label{prop:conditional_be_var}

Assume that mortality follows the affine Gaussian
model with deterministic coefficients, that discount
factors are deterministic, and that conditional
survival probabilities are computed consistently
with the underlying mortality dynamics and filtration.

For any projection date $i<T$, let
\begin{equation*}
s_i^2=V_i-V_{i+1}>0.
\end{equation*}

Then the logarithmic survival revision satisfies
\begin{equation*}
\mathcal R_i\mid\mathcal F_i
\sim
\mathcal N\left(
-\frac{s_i^2}{2},s_i^2
\right),
\end{equation*}
and the discounted Best Estimate variation admits
the representation
\begin{equation*}
\Delta_i
=
B_i\left(e^{\mathcal R_i}-1\right).
\end{equation*}

Consequently, the following properties hold:
\begin{equation*}
\begin{aligned}
\mathbb E[\mathcal R_i\mid\mathcal F_i]
&=-\frac{s_i^2}{2}<0,\\[2mm]
\mathbb E[\Delta_i\mid\mathcal F_i]
&=0,\\[2mm]
\mathbb P(\Delta_i<0\mid\mathcal F_i)
&=\Phi\left(\frac{s_i}{2}\right)>\frac12,
\end{aligned}
\end{equation*}
where $\Phi$ denotes the standard normal cumulative
distribution function.

Furthermore, for any confidence level $q\in(0,1)$,
the conditional Value-at-Risk of $\Delta_i$ is
\begin{equation}
\operatorname{VaR}_{q}(\Delta_i\mid\mathcal F_i)
=
B_i\left[
\exp\left(
-\frac{s_i^2}{2}
+s_i\Phi^{-1}(q)
\right)-1
\right].
\label{eq:analytical_var_general}
\end{equation}

In particular, the one-year capital requirement
at the $99.5\%$ confidence level is
\begin{equation}
SCR_i
=
B_i\left[
\exp\left(
-\frac{s_i^2}{2}
+s_i\Phi^{-1}(0.995)
\right)-1
\right].
\label{eq:analytical_scr_995}
\end{equation}

\end{proposition}

\begin{proof}

Since $Y_i=H_i+Y_{i+1}$, introduce
\begin{equation*}
Z_i=H_i+M_{i+1}.
\end{equation*}

As $H_i$ is $\mathcal F_{i+1}$-measurable,
the tower property of conditional expectations gives
\begin{align*}
\mathbb E[Z_i\mid\mathcal F_i]
&=
\mathbb E\left[
H_i+\mathbb E[Y_{i+1}\mid\mathcal F_{i+1}]
\mid\mathcal F_i
\right]\notag\\
&=
\mathbb E[Y_i\mid\mathcal F_i]
=M_i.
\end{align*}

Applying the law of total conditional variance
to $Y_i$ yields
\begin{align*}
V_i
&=
\mathbb E\left[
\operatorname{Var}(Y_i\mid\mathcal F_{i+1})
\mid\mathcal F_i
\right]\notag\\
&\quad+
\operatorname{Var}\left(
\mathbb E[Y_i\mid\mathcal F_{i+1}]
\mid\mathcal F_i
\right).
\end{align*}

Since $H_i$ is known at date $i+1$,
\begin{equation*}
\operatorname{Var}(Y_i\mid\mathcal F_{i+1})
=
V_{i+1},
\end{equation*}
and
\begin{equation*}
\mathbb E[Y_i\mid\mathcal F_{i+1}]
=
H_i+M_{i+1}
=
Z_i.
\end{equation*}

It follows that
\begin{equation*}
V_i
=
\mathbb E[V_{i+1}\mid\mathcal F_i]
+
\operatorname{Var}(Z_i\mid\mathcal F_i).
\end{equation*}

Under the affine Gaussian model with deterministic
coefficients, the conditional variance of integrated
mortality depends only on the projection dates
and model parameters, and not on the realized state.
Thus $V_{i+1}$ is deterministic, giving
\begin{equation*}
\operatorname{Var}(Z_i\mid\mathcal F_i)
=
V_i-V_{i+1}
=
s_i^2.
\end{equation*}

Moreover, $Z_i$ is affine in the Gaussian innovations
over $[i,i+1]$. Therefore,
\begin{equation*}
Z_i\mid\mathcal F_i
\sim
\mathcal N(M_i,s_i^2).
\end{equation*}

Using the closed-form Gaussian survival probability,
the two survival components can be expressed as
\begin{equation*}
D_i
=
\exp\left(
-M_i+\frac12V_i
\right)
\end{equation*}
and
\begin{equation*}
C_i
=
\exp\left(
-H_i-M_{i+1}+\frac12V_{i+1}
\right).
\end{equation*}

Consequently, the logarithmic survival revision is
\begin{align*}
\mathcal R_i
&=
\log\left(\frac{C_i}{D_i}\right)
\notag\\
&=
M_i-H_i-M_{i+1}
+\frac12(V_{i+1}-V_i)
\notag\\
&=
M_i-Z_i-\frac12s_i^2.
\end{align*}

Since $Z_i$ is conditionally Gaussian with mean
$M_i$ and variance $s_i^2$, we obtain
\begin{equation*}
\mathcal R_i\mid\mathcal F_i
\sim
\mathcal N\left(
-\frac{s_i^2}{2},s_i^2
\right).
\end{equation*}

In particular,
\begin{equation*}
\mathbb E[\mathcal R_i\mid\mathcal F_i]
=
-\frac{s_i^2}{2}<0.
\end{equation*}

Next, by
Proposition~\ref{prop:delta_i_decomposition},
the discounted Best Estimate variation is
\begin{align*}
\Delta_i
&=
P(i,T)\widetilde I(x,i)(C_i-D_i)
\notag\\
&=
P(i,T)\widetilde I(x,i)D_i
\left(e^{\mathcal R_i}-1\right)
\notag\\
&=
B_i\left(e^{\mathcal R_i}-1\right).
\end{align*}

Since $B_i$ is strictly positive and
$\mathcal F_i$-measurable,
\begin{align*}
\mathbb E[\Delta_i\mid\mathcal F_i]
&=
B_i\left(
\mathbb E[e^{\mathcal R_i}\mid\mathcal F_i]-1
\right).
\end{align*}

By the moment-generating function of the
normal distribution,
\begin{equation*}
\mathbb E[e^{\mathcal R_i}\mid\mathcal F_i]
=
\exp\left(
-\frac{s_i^2}{2}
+\frac12s_i^2
\right)
=1.
\end{equation*}

Hence,
\begin{equation*}
\mathbb E[\Delta_i\mid\mathcal F_i]=0.
\end{equation*}

Furthermore, because $B_i>0$,
\begin{equation*}
\Delta_i<0
\quad\Longleftrightarrow\quad
\mathcal R_i<0.
\end{equation*}

Using the conditional Gaussian distribution
of $\mathcal R_i$, we obtain
\begin{align*}
\mathbb P(\Delta_i<0\mid\mathcal F_i)
&=
\mathbb P(\mathcal R_i<0\mid\mathcal F_i)
\notag\\
&=
\Phi\left(\frac{s_i}{2}\right)
>\frac12.
\end{align*}

Finally, the function
$r\mapsto B_i(e^r-1)$ is strictly increasing.
The conditional quantile of $\Delta_i$ is therefore
obtained by applying this transformation to
the corresponding Gaussian quantile:
\begin{equation*}
\operatorname{VaR}_{q}(\Delta_i\mid\mathcal F_i)
=
B_i\left[
\exp\left(
-\frac{s_i^2}{2}
+s_i\Phi^{-1}(q)
\right)-1
\right].
\end{equation*}

Setting $q=0.995$ yields
equation~\eqref{eq:analytical_scr_995},
which completes the proof.

\end{proof}

\noindent\textbf{Projection of the conditional SCR.}
The initial Risk Margin in \eqref{eq:risk_margin_def} uses
$SCR_{i\mid0}=\mathbb E_0[SCR_i]$. Although $B_i$ in
\eqref{eq:analytical_scr_995} is random, its expectation is explicit.
Since $D_i={}_{T-i}p_{i,x+i}$, the tower property gives
\begin{align}
\mathbb E_0[B_i]
&=P(i,T)\,\mathbb E_0\!\left[
\widetilde I(x,i)\,{}_{T-i}p_{i,x+i}\right]\notag\\
&=P(i,T)\,\mathbb E_0\!\left[
\mathbb E[\widetilde I(x,T)\mid\mathcal F_i]\right]\notag\\
&=P(i,T)\,{}_{T}p_{0,x},
\label{eq:projected_expected_b}
\end{align}
where ${}_{T}p_{0,x}=\mathbb E_0[\widetilde I(x,T)]$
is the model-implied survival probability at inception.
As $s_i^2=V_i-V_{i+1}$ and the discount factors are
deterministic, the projected capital requirement is therefore
\begin{equation}
\label{eq:projected_scr_analytic}
SCR_{i\mid0}
=P(i,T)\,{}_{T}p_{0,x}
\left[\exp\left(
-\frac{s_i^2}{2}+s_i\Phi^{-1}(0.995)
\right)-1\right].
\end{equation}
This is the analytical expected conditional SCR used in the
Risk Margin calculations reported below.

The proposition shows that the logarithmic survival
revision has a negative conditional expectation,
although the discounted Best Estimate variation
has zero conditional expectation. Negative variations
are therefore more frequent, while less frequent
positive variations compensate for them in expectation.
This asymmetry is consistent with a positive upper
quantile and provides an analytical characterization
of the one-year Value-at-Risk under the affine
Gaussian mortality model.


\section{Stress-based assessment of structural longevity risk}
\label{sec:structural_stress}

The Solvency II-consistent pricing framework developed in Section~\ref{sec:solvency_pricing} evaluates longevity risk through the stochastic revaluation of future survival probabilities and the associated one-year capital requirements. This approach provides a distribution-based measure of adverse liability variations. However, structural longevity risk also arises from the possible underestimation of persistent mortality improvements. Such risk is not only a matter of short-term stochastic volatility, but also of adverse revisions in the long-term assumptions governing the mortality trend.\\

Stress-based valuation provides a complementary actuarial perspective by measuring the sensitivity of longevity-linked liabilities to deterministic adverse changes in the mortality dynamics. In the present framework, two stress specifications are considered. The first is a mortality-level stress applied directly to the mortality intensity. The second is a trend-based longevity stress applied to the deterministic target of the latent longevity factor. The distinction is important because the two stresses affect the affine mortality model through different channels. The mortality-level stress modifies the current and future intensity surface proportionally, whereas the trend-based stress modifies the prospective dynamics of long-term mortality improvements.

\subsection{Mortality-level stress}
\label{subsec:mortality_level_stress}

The mortality-level stress is defined as an instantaneous and permanent reduction in mortality intensities. This specification is in line with the actuarial interpretation of a longevity shock as a decrease in mortality rates, which leads to higher survival probabilities and larger values of longevity-sensitive liabilities. For a stress parameter $K\in(0,1)$, the stressed mortality intensity is given by
\begin{equation}
	\label{eq:level_stressed_mu}
	\mu^{\mathrm{lev}}_{t,x}
	=
	(1-K)\mu_{t,x}.
\end{equation}
A positive value of $K$ therefore corresponds to an adverse longevity scenario, since lower mortality intensities imply higher survival probabilities and, consequently, larger liability values.\\

For each projection date $i<T$, the stress is applied prospectively, conditional on the mortality state and cumulative survival observed under the central model at date $i$. The conditional survival probability is obtained by replacing the future mortality intensity by its stressed counterpart:
\begin{equation*}
 {}_{T-i}p^{\mathrm{lev}}_{i,x+i}
 = \mathbb E\left[\exp\left(-\int_0^{T-i}\mu^{\mathrm{lev}}_{i+u,x+i+u}\,du\right)\middle|\mathcal F_i\right].
\end{equation*}

The following proposition records the closed-form implication of this proportional stress under the Gaussian affine survival representation derived from Hainaut (2023)~\cite{Hainaut2023}.\\

\begin{proposition}[Mortality-level stress under affine Gaussian survival]
 \label{prop:level_stress_survival}
 Assume that, conditionally on $\mathcal F_i$, the integrated central mortality intensity over $[i,T]$ is Gaussian with mean ${}_{T-i}m_{i,x+i}$ and variance ${}_{T-i}v_{i,x+i}$. Under the mortality-level stress \eqref{eq:level_stressed_mu}, the stressed conditional survival probability is
 \begin{equation}
 \label{eq:level_stressed_survival_closed}
 {}_{T-i}p^{\mathrm{lev}}_{i,x+i}
 =\exp\left(-(1-K)\,{}_{T-i}m_{i,x+i}
 +\frac12(1-K)^2\,{}_{T-i}v_{i,x+i}\right).
 \end{equation}
\end{proposition}

\begin{proof}
 The integrated future stressed intensity equals $(1-K)\int_i^T\mu_{t,x+t}\,dt$. Its conditional mean and variance are therefore respectively $(1-K)\,{}_{T-i}m_{i,x+i}$ and $(1-K)^2\,{}_{T-i}v_{i,x+i}$. The Gaussian moment-generating function yields the result.
\end{proof}

This stress acts directly on prospective mortality intensities while preserving the already observed central mortality state. It provides a transparent benchmark against which the structural stress introduced next can be compared.

\subsection{Trend-based longevity stress}
\label{subsec:trend_based_stress}

The trend-based longevity stress differs from the mortality-level stress because it does not directly rescale the mortality intensity. Instead, it modifies the deterministic target governing the latent longevity factor. This specification is designed to capture adverse revisions in long-term mortality improvement assumptions.\\

Recall that the latent longevity factor follows \eqref{eq:theta_dyn}, with the deterministic reversion target defined in \eqref{eq:beta_def}.
The trend stress is introduced through the parameters governing this long-term target:
\begin{equation}
	\label{eq:trend_stressed_parameters}
	\theta_0^{\mathrm{tr}}
	=
	(1-K)\theta_0,
	\qquad
	\bar\theta^{\mathrm{tr}}
	=
	(1-K)\bar\theta.
\end{equation}
Consequently, the stressed reversion target becomes
\begin{equation*}
	\beta^{\mathrm{tr}}(t)
	=
	(1-K)\beta(t),
\end{equation*}
and the stressed longevity factor follows
\begin{equation}
	\label{eq:theta_trend_dynamics}
	d\theta_t^{\mathrm{tr}}
	=
	\alpha
	\left(
	\beta^{\mathrm{tr}}(t)
	-
	\theta_t^{\mathrm{tr}}
	\right)dt
	+
	\nu\,dW_t^\theta.
\end{equation}

The stressed mortality intensity is then obtained by replacing the central longevity factor with the stressed one in the mortality dynamics. The stress therefore affects mortality through the structural component governing long-term improvements, rather than through a direct proportional change in the mortality level.\\

At each projection date $i$, the stress is applied only to the future target of the latent factor. The current mortality curve and $\theta_i$ remain at their central simulated values: $\theta_i^{\mathrm{tr}}=\theta_i^{\mathrm{central}}$ when the stressed projection is initiated. Conditional on $\mathcal F_i$, the stressed remaining survival probability is
\begin{equation}
 \label{eq:trend_stressed_survival_closed}
 {}_{T-i}p^{\mathrm{tr}}_{i,x+i}
 =\exp\left(-{}_{T-i}m^{\mathrm{tr}}_{i,x+i}
 +\frac12\,{}_{T-i}v^{\mathrm{tr}}_{i,x+i}\right),
\end{equation}
where the conditional moments are evaluated under the future stressed trend target, starting from the central state at $i$. The volatility parameters are held fixed; the scenario therefore changes the projected conditional mean rather than imposing an instantaneous shock on the past state.

The Best Estimate under the trend stress at the \emph{same valuation date} $i$ is consequently
\begin{equation}
 \label{eq:be_trend_stressed}
 BE_i^{\mathrm{tr}}
 =P(i,T)\left(
 \widetilde I^{\mathrm{central}}(x,i)\,
 {}_{T-i}p^{\mathrm{tr}}_{i,x+i}
 -{}_{T-t_0}\bar p_x
 \right).
\end{equation}
This construction preserves the central cumulative survival up to $i$ and applies the revised dynamics only over the remaining horizon. The mortality-level stress is evaluated using the analogous same-date formula with ${}_{T-i}p^{\mathrm{lev}}_{i,x+i}$.

\subsection{Scenario-based capital requirements and Risk Margin}
\label{subsec:stress_valuation_adjustment}

The two scenarios are applied to the \emph{future} mortality dynamics, while
keeping the state observed under the central model at the valuation date
unchanged. Table~\ref{tab:stress_comparison} summarizes their distinct
transmission mechanisms.

\begin{table}[htbp]
 \centering
 \caption{Comparison of mortality-level and trend-based longevity stresses.}
 \label{tab:stress_comparison}
 \renewcommand{\arraystretch}{1.3}
 \begin{tabular}{@{}p{4.0cm}p{4.8cm}p{4.8cm}@{}}
 \toprule
 & \textbf{Mortality-level stress} & \textbf{Trend-based stress} \\
 \midrule
 \textbf{Stress variable} & $\mu_{t,x}$ & $\beta(t)$ \\
 \textbf{Transformation} & $\mu^{\mathrm{lev}}_{t,x}=(1-K)\mu_{t,x}$ & $\beta^{\mathrm{tr}}(t)=(1-K)\beta(t)$ \\
 \textbf{Effect on survival} & Direct rescaling of prospective mortality intensities & Change in projected longevity-factor dynamics \\
 \textbf{Dynamic effect} & Proportional mortality-level shock & Change in the prospective mean-reversion target \\
 \textbf{Actuarial interpretation} & Adverse reduction in future mortality rates & Structural revision of future mortality improvements \\
 \bottomrule
 \end{tabular}
\end{table}
\FloatBarrier

\medskip
\noindent\textbf{Same-date valuation.}

For each valuation date $i<T$, let $s\in\{\mathrm{lev},\mathrm{tr}\}$
denote the stress type and $K\in\{0.10,0.20,0.30\}$ its intensity.
Conditional on the central history $\mathcal F_i$, the stressed Best
Estimate is evaluated at the same date $i$ as its central counterpart:
\begin{equation}
\label{eq:stress_be_same_date}
BE_i^{s,K}=P(i,T)\left(
\widetilde I^{\mathrm{central}}(x,i)\,
{}_{T-i}p^{s,K}_{i,x+i}
-{}_{T-t_0}\bar p_x
\right).
\end{equation}

Only the future mortality projection is stressed; the survival history
realized up to $i$ is left unchanged.

\medskip
\noindent\textbf{Scenario-based capital requirement.}

We define the projected capital requirement as the positive part of
the expected increase in the Best Estimate:
\begin{equation}
\label{eq:stress_projected_scr}
\mathrm{SCR}_{i|0}^{s,K,\mathrm{stress}}
=\left[\mathbb E_0\!\left(
BE_i^{s,K}-BE_i^{\mathrm{central}}
\right)\right]_+,
\qquad [z]_+=\max(z,0).
\end{equation}

In the Monte Carlo implementation, valuation differences are averaged
across central paths \emph{before} applying the positive part.

\medskip
\noindent\textbf{Cost-of-Capital aggregation.}

The scenario-based Risk Margin is obtained by aggregating the
projected capital requirements:
\begin{equation}
\label{eq:stress_rm_same_date}
\mathrm{RM}_0^{s,K,\mathrm{stress}}
=c_{\mathrm{CoC}}\sum_{i=0}^{T-1}
P(0,i+1)\,\mathrm{SCR}_{i|0}^{s,K,\mathrm{stress}}.
\end{equation}

\Needspace{7\baselineskip}

\subsection{Algorithmic implementation}
\label{subsec:algorithmic_implementation}

Algorithm~\ref{alg:trend_stress_be} summarizes the trend-stress valuation
at an arbitrary projection date $i<T$. As in the original construction,
the stress modifies the future mortality dynamics. The key adjustment is
that the conditional survival is projected over $[i,T]$ from the central
state observed at $i$, without re-simulating the survival history before
$i$. Both Best Estimates are therefore evaluated at the same date.

\begin{algorithm}[htbp]
 \caption{Trend-stress valuation and projected capital at date $i$}
 \label{alg:trend_stress_be}
 \begin{algorithmic}[1]
  \Require Central states $(\theta_i^{(n)},\mu_{i,x+i}^{(n)})$ and cumulative survival
  $\widetilde I^{\mathrm{central},(n)}(x,i)$ for $n=1,\ldots,N$; stress level $K$;
  maturity $T$ and central mortality parameters.
  \Ensure Projected trend-stress capital $\mathrm{SCR}_{i|0}^{\mathrm{tr},K,\mathrm{stress}}$.
  \Statex \vspace{0.75em}
  \State Construct the stressed future trend target:
  \Statex \hspace{\algorithmicindent}$\displaystyle \beta^{\mathrm{tr}}(t)=(1-K)\beta(t),\qquad t\geq i.$
  \Statex \vspace{0.75em}
  \State Keep the central state at date $i$ unchanged:
  \Statex \hspace{\algorithmicindent}$\displaystyle
   \theta_i^{\mathrm{tr},(n)}=\theta_i^{\mathrm{central},(n)},\qquad
   \mu_{i,x+i}^{\mathrm{tr},(n)}=\mu_{i,x+i}^{\mathrm{central},(n)}.$
  \Statex \vspace{0.75em}
  \State Evaluate the conditional moments of future integrated mortality
  under the stressed dynamics over $[i,T]$.
  \Statex \vspace{0.75em}
  \State Compute the stressed remaining survival probability:
  \Statex \hspace{\algorithmicindent}$\displaystyle
   {}_{T-i}p_{i,x+i}^{\mathrm{tr},K,(n)}
   =\exp\!\left(-{}_{T-i}m_{i,x+i}^{\mathrm{tr},K,(n)}
   +\frac12\,{}_{T-i}v_{i,x+i}^{\mathrm{tr},K,(n)}\right).$
  \Statex \vspace{0.75em}
  \State Deduce the stressed Best Estimate at date $i$:
  \Statex \hspace{\algorithmicindent}$\displaystyle
   BE_i^{\mathrm{tr},K,(n)}=P(i,T)\left(
   \widetilde I^{\mathrm{central},(n)}(x,i)
   {}_{T-i}p_{i,x+i}^{\mathrm{tr},K,(n)}
   -{}_{T-t_0}\bar p_x\right).$
  \Statex \vspace{0.75em}
  \State Evaluate $BE_i^{\mathrm{central},(n)}$ at the same date, using
  the central conditional survival probability.
  \Statex \vspace{0.75em}
  \State Average the adverse revaluation and apply the positive part:
  \Statex \hspace{\algorithmicindent}$\displaystyle
   \mathrm{SCR}_{i|0}^{\mathrm{tr},K,\mathrm{stress}}
   =\max\!\left\{0,\,\frac1N\sum_{n=1}^{N}
   \big(BE_i^{\mathrm{tr},K,(n)}-BE_i^{\mathrm{central},(n)}\big)\right\}.$
  \Statex \vspace{0.75em}
  \State \Return $\mathrm{SCR}_{i|0}^{\mathrm{tr},K,\mathrm{stress}}$.
 \end{algorithmic}
\end{algorithm}
\FloatBarrier

The procedure is repeated for $i=0,\ldots,T-1$ and the projected
capital terms are aggregated using \eqref{eq:stress_rm_same_date}.
For the mortality-level stress, the same-date valuation principle applies,
with the future mortality intensity rescaled by $(1-K)$ instead of modifying
$\beta(t)$. The empirical results are presented in
Subsection~\ref{subsec:stress_pricing_results}.


\section{Empirical application to Belgian mortality data}
\label{sec:empirical_application}

This section illustrates the implementation of the proposed Solvency II-consistent pricing framework using Belgian mortality data. The empirical analysis has three objectives. First, it assesses whether the affine mortality model provides a sufficiently stable representation of the observed mortality surface. Second, it evaluates the resulting Best Estimate, one-year SCR and Risk Margin under the baseline stochastic mortality dynamics. Third, it compares the VaR-based capital assessment with the scenario-based capital approach developed in Section~\ref{sec:structural_stress}.

\subsection{Data and calibration procedure}
\label{subsec:data_calibration}

The model is calibrated using mortality data for the Belgian population over the period 1970--2022, obtained from the Human Mortality Database \cite{HMD}. The calibration uses ages 0–109, while graphical results are presented over the restricted age range 0–100 for visualization purposes. The dataset provides the death counts and exposures required to estimate age-specific mortality intensities. The years 2023 and 2024 are excluded from the calibration window and used for out-of-sample validation in Subsection~\ref{subsec:out_of_sample_validation}.\\

\begin{figure}[htbp]
	\centering
	\begin{minipage}[t]{0.48\textwidth}
		\centering
		\includegraphics[height=4.2cm]{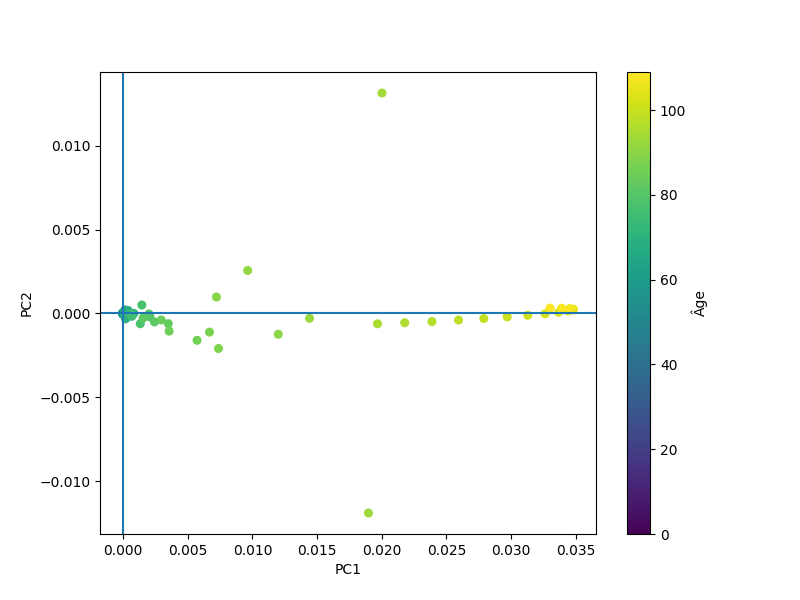}\\[-0.2em]
		{\footnotesize (a) Principal component representation}
	\end{minipage}
	\hfill
	\begin{minipage}[t]{0.48\textwidth}
		\centering
		\includegraphics[height=4.2cm]{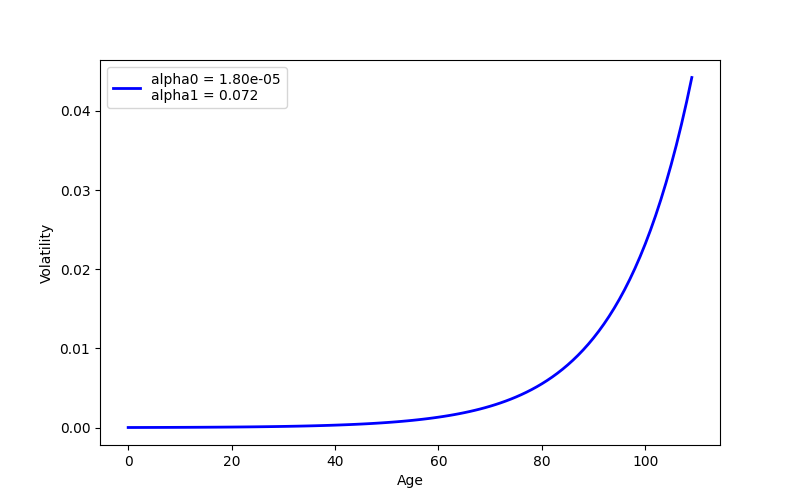}\\[-0.2em]
		{\footnotesize (b) Exponential volatility function}
	\end{minipage}
	\caption{Mortality variability and fitted volatility specification.}
	\label{fig:pca_volatility}
\end{figure}

The calibration is carried out in two steps. First, a principal component analysis is performed on historical mortality variations in order to identify the dominant age pattern of mortality volatility. Figure~\ref{fig:pca_volatility} reports the empirical principal component representation and the fitted volatility function. The dispersion remains limited at younger ages and becomes more pronounced at advanced ages, especially along the dominant component. From an actuarial perspective, this pattern is consistent with the higher instability of mortality estimates at older ages, where exposures are smaller and observed death rates are more sensitive to sampling variability.\\

This empirical structure supports a parsimonious one-factor specification for the age-dependent volatility function. We therefore adopt the exponential form

\begin{equation*}
    \Sigma(x)=\sigma_0 e^{\sigma_1 x}.
\end{equation*}

This specification captures the increasing uncertainty of mortality at older ages while preserving the analytical tractability of the affine mortality framework. In particular, it remains compatible with the closed-form survival probabilities used in the Best Estimate, SCR and Risk Margin computations. Additional numerical details on the calibration of the volatility function are reported in Appendix~\ref{app:volatility_calibration}.\\

The estimated parameters of the volatility function are reported in 
Table~\ref{tab:params_pca}. 
These estimates are preliminary values obtained from the empirical/PCA step 
and are subsequently used as initial values for the global calibration, 
in which the volatility parameters $\sigma_0$ and $\sigma_1$ are re-estimated 
jointly with the other model parameters.
The positive value of $\widehat{\sigma}^{\,\mathrm{PCA}}_1$ confirms the 
increasing age profile of mortality volatility. This feature is important 
for the subsequent valuation exercise, because longevity-linked payoffs 
and solvency capital requirements are sensitive to the uncertainty embedded 
in future survival probabilities.

\begin{table}[htbp]
	\centering
	\caption{Estimated volatility parameters.}
	\label{tab:params_pca}
	\begin{tabular}{cc}
		\toprule
		\textbf{Parameter} & \textbf{Estimated value} \\
		\midrule
		$\widehat{\sigma}^{\,\mathrm{PCA}}_0$ 
		& $2.42 \times 10^{-5}$ \\
		$\widehat{\sigma}^{\,\mathrm{PCA}}_1$ 
		& $0.068$ \\
		\bottomrule
	\end{tabular}
\end{table}

In the second step, the full set of model parameters is estimated by maximum likelihood using the conditional Gaussian representation associated with the affine mortality framework of Hainaut (2023)\cite{Hainaut2023}. Since the longevity factor is latent, likelihood evaluation is combined with a recursive filtering procedure that updates the conditional distribution of the hidden factor from observed mortality intensities. The resulting parameter estimates are reported in Table~\ref{tab:params_classique}.

\begin{table}[htbp]
    \centering
    \caption{Estimated model parameters.}
    \label{tab:params_classique}
    \begin{tabular}{cccc}
        \toprule
        \textbf{Parameter} & \textbf{Estimated value}
        & \textbf{Parameter} & \textbf{Estimated value} \\
        \midrule
        $\kappa$   & $4.986\times10^{-2}$
        & $\theta_0$ & $4.149\times10^{-1}$ \\

        $\alpha$   & $5.000\times10^{1}$
        & $\bar{\theta}$ & $1.933\times10^{-1}$ \\

        $\gamma$   & $9.452\times10^{-1}$
        & $\nu$ & $1.053\times10^{1}$ \\

        $\sigma_0$ & $6.664\times10^{-5}$
        & $\sigma_1$ & $4.869\times10^{-2}$ \\

        $a$ & $1.001\times10^{-8}$
        & $b$ & $1.355\times10^{-4}$ \\

        $c$ & $1.101$ & & \\
        \bottomrule
    \end{tabular}

    \vspace{0.5em}

    \begin{minipage}{0.95\linewidth}
        \footnotesize
        \color{green!50!black}      
    \end{minipage}
\end{table}

The parameter estimates provide the numerical basis for the survival probabilities and valuation quantities computed below. The Gompertz--Makeham parameters determine the baseline age profile of mortality, while the parameters $\theta_0$, $\bar{\theta}$ and $\gamma$ govern the deterministic target of the latent longevity factor. These latter parameters are particularly important in the present framework, because they control the long-term mortality improvement component on which the trend-based stress is applied. An illustration of the closed-form survival probability calculation is provided in Appendix~\ref{app:closed_form_survival}.\\

\begin{figure}[htbp!]
    \centering

    \begin{subfigure}[t]{0.49\textwidth}
        \centering
        \includegraphics[width=\linewidth]
        {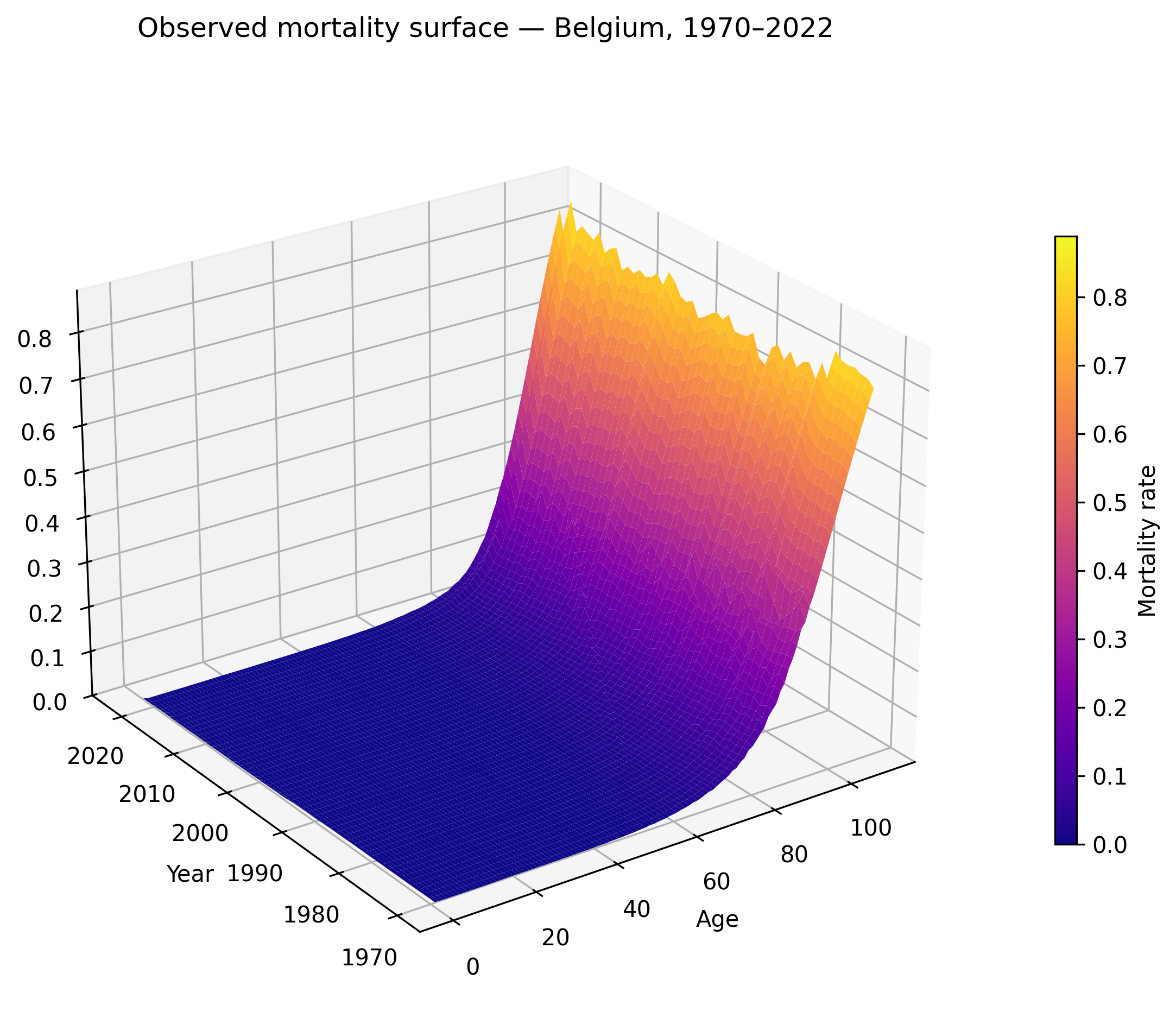}
        \caption{Observed mortality surface, 1970--2022.}
        \label{fig:observed_mortality_surface}
    \end{subfigure}
    \hfill
    \begin{subfigure}[t]{0.49\textwidth}
        \centering
        \includegraphics[width=\linewidth]
        {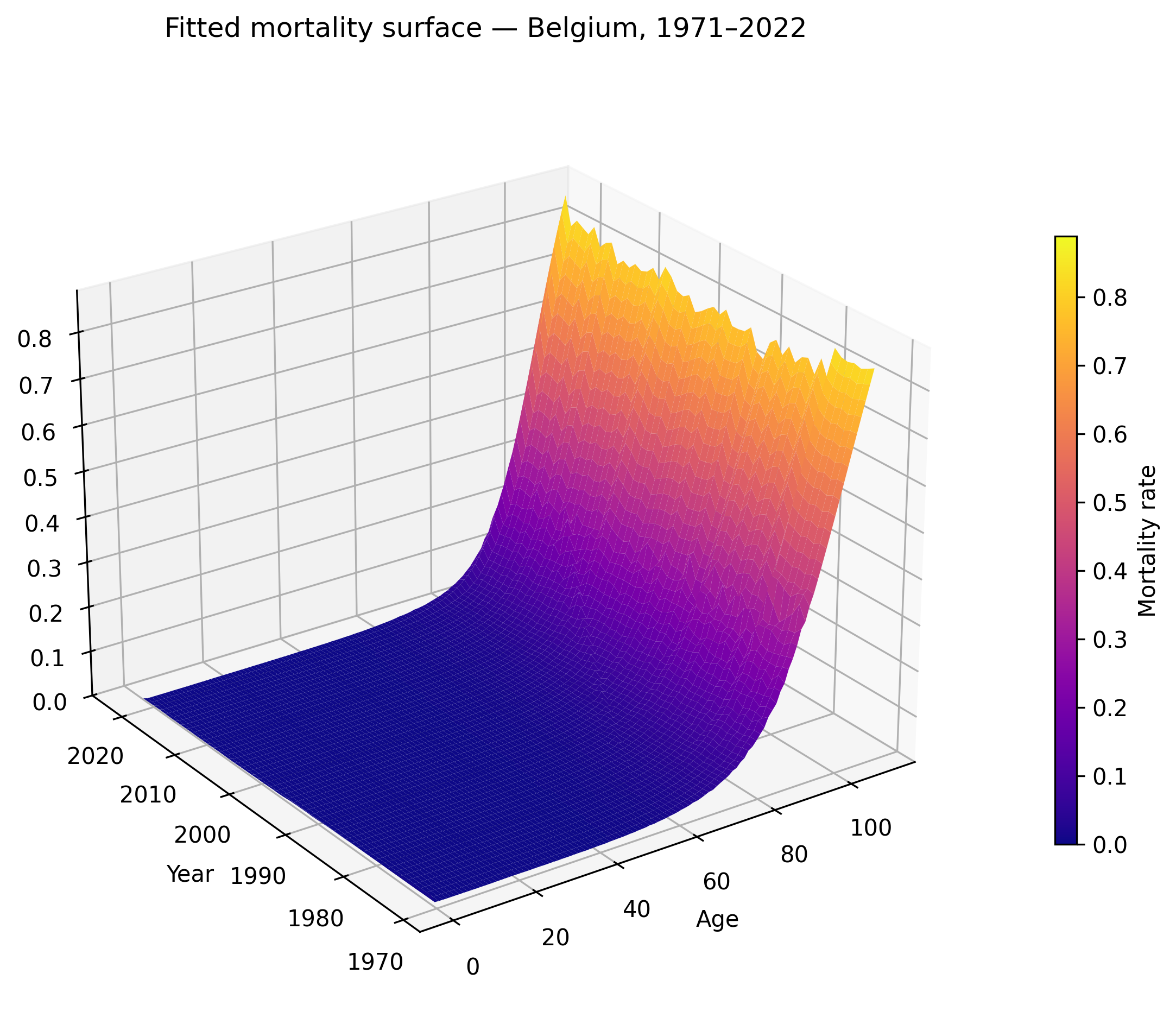}
        \caption{Fitted mortality surface, 1971--2022.}
        \label{fig:fitted_mortality_surface}
    \end{subfigure}

    \caption{Observed and fitted mortality surfaces for Belgium.
    For each year $t$, the fitted rates correspond to the
    one-year-ahead conditional means
    $\widehat{\boldsymbol{\mu}}_{t\mid t-1}
    =
    m_\mu(t-1,
    \boldsymbol{\mu}^{\mathrm{obs}}_{t-1},
    \widehat{\theta}_{t-1})$.
    The fitted surface is conditional on the observed
    mortality intensities and the filtered longevity
    factor from the preceding year.}
    \label{fig:observed_fitted_mortality_surfaces}
\end{figure}

To complement the parameter estimates, the calibration fit
is assessed by comparing the observed and fitted mortality
surfaces for Belgium, as shown in
Figure~\ref{fig:observed_fitted_mortality_surfaces}.
The observed surface covers the calibration period
1970--2022, whereas the fitted surface covers
1971--2022 and represents the one-year-ahead conditional
mean mortality intensities obtained from the calibrated
model. The comparison provides a visual assessment of
the model's ability to reproduce the age-specific
mortality structure and its evolution over time.

The observed and fitted mortality surfaces exhibit
similar age-dependent patterns and temporal evolution.
In particular, the fitted surface reproduces the
pronounced increase in mortality at advanced ages.
This visual agreement is consistent with the temporal
persistence embedded in the model, whereby the
conditional mean depends on the mortality intensities
observed in the preceding year. Consequently, the
similarity between the two surfaces should not be
interpreted as sufficient evidence of predictive
accuracy on its own. The model's forecasting performance
is assessed separately through the out-of-sample
validation for 2023 and 2024.

These diagnostics are computed for the Belgian population over 1970--2022, with ages 0--100 displayed for graphical readability. The calibration itself uses ages 0--109, as described above. The observed surface displays the expected increase in mortality intensity with age, while the residual heatmap provides a diagnostic view of the deviations between observed and fitted mortality intensities. The residuals remain close to zero for most ages and do not exhibit pronounced temporal patterns over the central part of the age distribution. Larger deviations are mainly concentrated at advanced ages, where exposures are lower and mortality estimates are inherently more volatile. This behaviour is consistent with the volatility pattern identified above and does not materially affect the valuation setting considered below, which focuses on maturities and portfolio ages below the most unstable part of the mortality surface.\\


\subsection{Out-of-sample validation}
\label{subsec:out_of_sample_validation}

The predictive performance of the calibrated affine mortality
model is assessed over 2023 and 2024, which were excluded
from the calibration period. The validation compares observed
and projected mortality intensities using the exposure-weighted
error measures, pointwise predictive interval coverage and
standardized Gaussian predictive residuals introduced in
Subsection~\ref{subsec:observation_validation}.
This out-of-sample assessment is particularly relevant
because the subsequent valuation framework relies on the
ability of the mortality model to generate prospective
survival probabilities.

Table~\ref{tab:validation} reports the validation results.
Exposure-weighted metrics are particularly relevant in an
actuarial setting, as prediction errors at ages with larger
exposures receive greater weight than errors at sparsely
populated ages.

\begin{table}[!htbp]
    \centering
    \caption{Out-of-sample validation results.}
    \label{tab:validation}
    \begin{tabular}{cccc}
        \toprule
        \textbf{Year}
        & $\mathbf{MAE}_{\mathrm{w}}$
        & $\mathbf{RMSE}_{\mathrm{w}}$
        & \textbf{95\% coverage} \\
        \midrule
        2023
        & $7.705\times10^{-4}$
        & $3.757\times10^{-3}$
        & $82.7\%$ \\
        2024
        & $8.977\times10^{-4}$
        & $2.523\times10^{-3}$
        & $99.1\%$ \\
        \bottomrule
    \end{tabular}
\end{table}

The exposure-weighted errors remain relatively small in
absolute terms over the two validation years, although
their evolution is not uniform. The weighted MAE increases
slightly from 2023 to 2024, whereas the weighted RMSE
decreases. This suggests that the magnitude and age
distribution of prediction errors differ between the
two years.

The empirical coverage of the pointwise 95\% predictive
intervals reaches 82.7\% in 2023 and 99.1\% in 2024.
The lower coverage in 2023 indicates that the predictive
intervals do not fully capture the observed age-specific
mortality variations in that year. Conversely, the high
coverage in 2024 suggests that the simulated predictive
distribution encompasses most observed mortality rates.
However, coverage above the nominal level may also
reflect relatively wide predictive intervals and should
not be interpreted as evidence of superior point
forecast accuracy.

\begin{figure}[!htbp]
    \centering
    \includegraphics[width=0.92\textwidth]
    {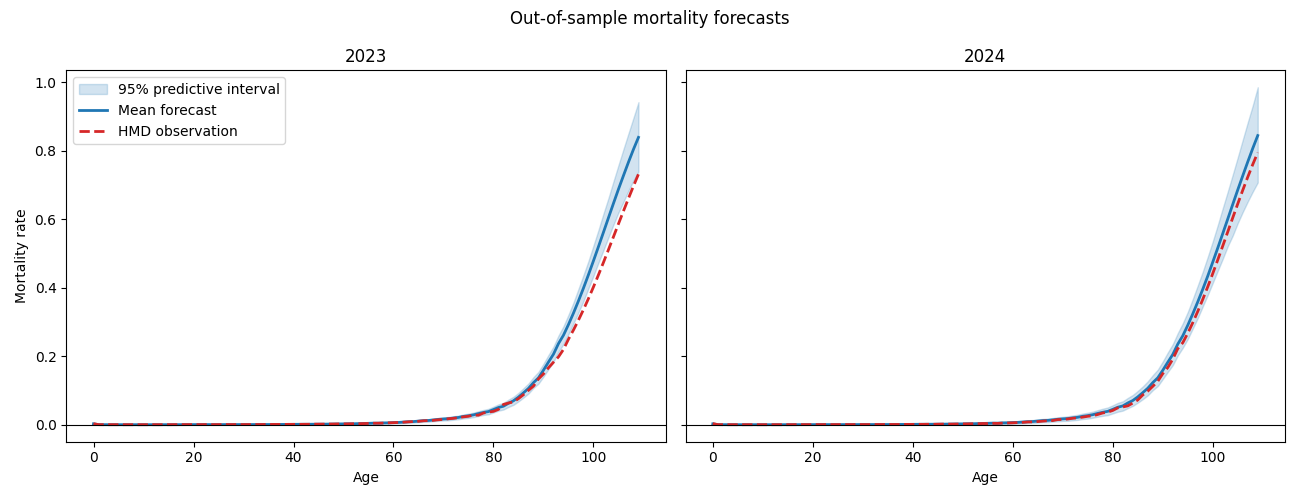}
    \caption{Out-of-sample mortality forecasts for
    Belgium in 2023 and 2024. Solid blue curves
    represent the simulated mean forecasts, shaded
    areas denote the pointwise 95\% predictive
    intervals, and dashed red curves represent
    the observed HMD mortality rates.}
    \label{fig:mortality_forecast}
\end{figure}

Figure~\ref{fig:mortality_forecast} compares the
observed mortality curves with the simulated mean
forecasts and the corresponding pointwise 95\%
predictive intervals for both validation years.

The forecast curves reproduce the overall age-dependent shape of observed mortality, particularly the pronounced increase at advanced ages. Nevertheless, deviations
remain visible, indicating that the model does not capture all age-specific mortality variations.
The predictive intervals provide complementary
information on the dispersion of the simulated
mortality intensities.

\begin{figure}[!htbp]
    \centering
    \includegraphics[width=0.88\textwidth]
    {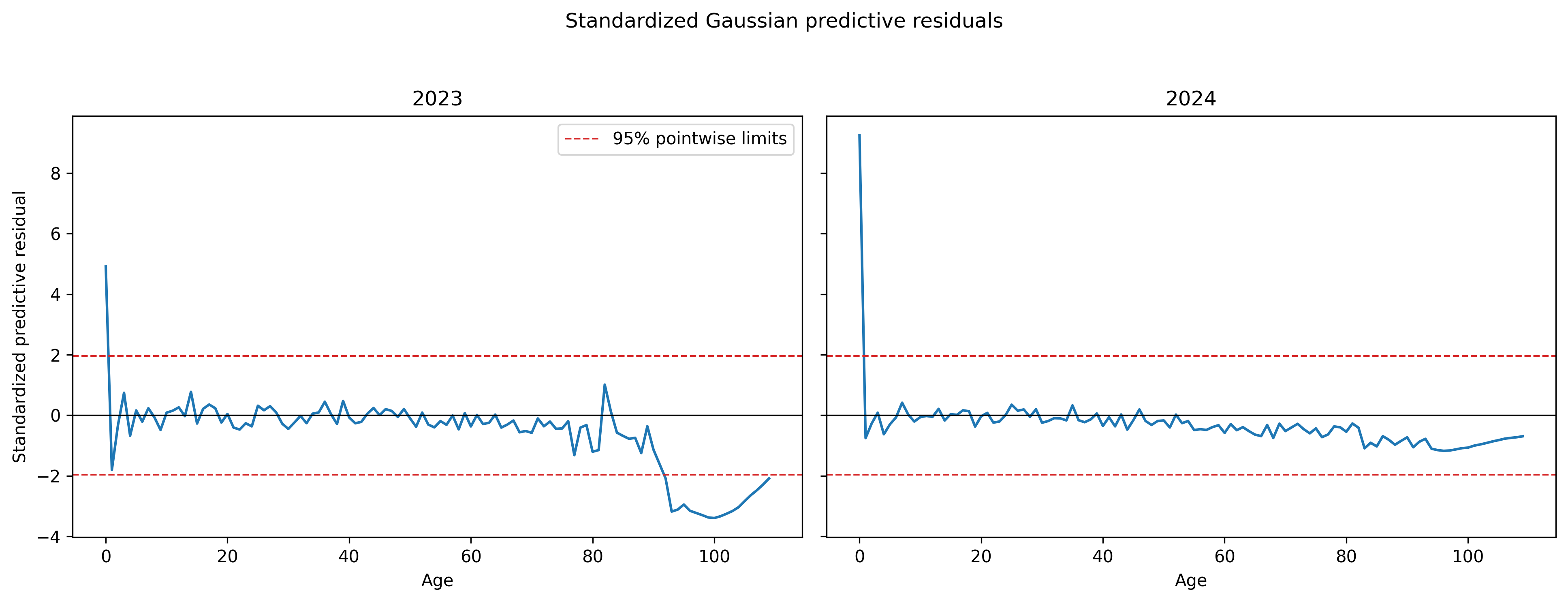}
    \caption{Standardized Gaussian predictive residuals
    for Belgium in 2023 and 2024. The dashed horizontal
    lines indicate the pointwise reference bounds
    of $\pm 1.96$.}
    \label{fig:residuals}
\end{figure}

To examine these deviations more closely,
Figure~\ref{fig:residuals} presents standardized
Gaussian predictive residuals by age for 2023 and
2024. The horizontal dashed lines indicate the
pointwise Gaussian reference bounds of $\pm 1.96$.

The residuals are relatively concentrated around zero
over a substantial part of the age distribution.
However, pronounced positive deviations are observed
at the youngest ages, particularly in 2024, while
negative deviations occur at advanced ages in 2023.
Several residuals exceed the pointwise reference
bounds, indicating localized discrepancies between
observed mortality intensities and the predictive
distribution.

These deviations highlight limitations in the
model's ability to reproduce mortality variations
uniformly across all ages. Moreover, the residuals
are marginal diagnostics and do not account directly
for the dependence structure across ages.

Overall, the validation results indicate that the
calibrated affine mortality model captures the broad
age-dependent mortality structure and provides
a tractable framework for short-horizon forecasting.
Nevertheless, the differences in predictive coverage
between 2023 and 2024 and the localized residual
deviations suggest that predictive uncertainty is
not uniformly calibrated across ages and years.

These limitations should be considered when using
the model for the subsequent valuation and capital
assessment exercises. Its analytical tractability,
particularly the availability of closed-form
survival probabilities, remains advantageous for
repeated Best Estimate, SCR and Risk Margin
calculations under stochastic and stressed
mortality scenarios.

\FloatBarrier

\needspace{8\baselineskip}
\subsection{VaR-based capital assessment}
\label{subsec:var_capital_assessment}

The illustrative portfolio consists of $10{,}000$ homogeneous longevity-linked
contracts, with an initial age of 60 and a 15-year horizon. We use a flat
risk-free rate of $1\%$ and a Cost-of-Capital rate of $6\%$, consistent with
the Solvency II provisions applicable at the 2023 valuation date under
Commission Delegated Regulation (EU) 2015/35 \cite{EUDelegated2015}.
The revised 2027 aggregation is examined in Appendix~\ref{app:nv_reforme}
\cite{EUDelegated2026}.

The fixed survival benchmark is obtained from the 2023 Belgian period life
table (Human Mortality Database, Life Tables 1x1):
\[
{}_{15}\bar p_{60}
=\prod_{a=60}^{74}(1-q_a^{\mathrm{HMD},2023})
=0.824417.
\]
It is held unchanged across all central and stressed valuations.\\

At inception, the model-implied 15-year survival
probability is 0.7911, compared with 0.8244 for
the fixed 2023 period-table benchmark. To assess
the sensitivity to age interpolation, the 2023
annual death probabilities are converted into
intensities, $\mu_x=-\log(1-q_x)$, and linearly
interpolated across ages. This yields a static
continuous-age survival probability of 0.8167,
a difference of 0.77 percentage points relative
to the original benchmark.

The additional discrepancy with the model-implied
survival probability reflects prospective mortality
dynamics and differences in the information sets.
In particular, the estimated long-run mortality
target exceeds recent observed intensities at
representative ages. Moreover, the model is
calibrated using data through 2022, whereas the
benchmark relies on the 2023 period table.
Under deterministic discounting, the fixed
benchmark cancels from one-year discounted
Best Estimate variations and therefore does
not directly affect the analytical SCRs or
the associated Risk Margin.
\\

At each projection date, the Best Estimate is updated using the stochastic survival index and the remaining conditional survival probability. The one-year variation is defined in \eqref{eq:delta_i_def}. The primary capital
measure is the expected conditional VaR obtained from
\eqref{eq:analytical_scr_995} and aggregated using \eqref{eq:risk_margin_def}.
For comparison, we estimate the same expected conditional capital by Monte Carlo.

\begin{figure}[!htbp]
	\centering
	\includegraphics[width=0.67\textwidth]{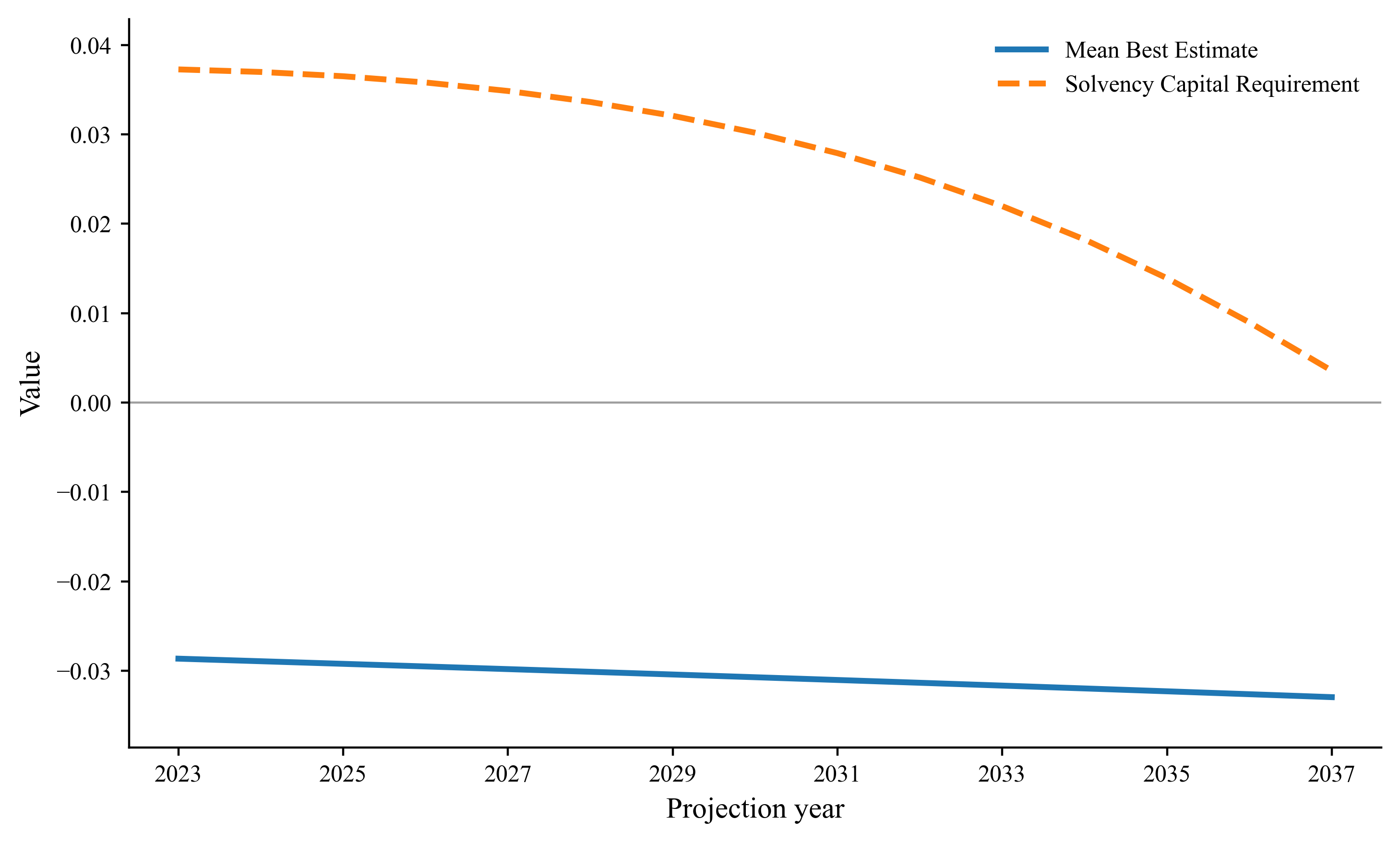}
	\caption{Evolution of the mean Best Estimate and the solvency capital requirement.}
	\label{fig:be_scr_relative}
\end{figure}

Figure~\ref{fig:be_scr_relative} presents the analytical expected conditional capital profile alongside the mean Best Estimate. The mean Best Estimate remains negative, changing from approximately $-0.0286$ in 2023 to $-0.0329$
in 2037, while the expected conditional SCR decreases smoothly from $0.0373$ to $0.0036$. This decline reflects the shortening of the remaining exposure horizon and the associated reduction in one-year longevity risk.

Table~\ref{tab:rm_summary} reports the reference analytical Risk Margin, based on the expected conditional VaR and a $6\%$ Cost-of-Capital rate.

\begin{table}[!htbp]
\centering
\caption{Analytical conditional VaR-based Risk Margin for the reference portfolio.}
\label{tab:rm_summary}
\begin{tabular}{lc}
\toprule
Indicator & Value \\
\midrule
$RM_0$ (unit) & $0.0223664$ \\
Portfolio size & $10{,}000$ \\
$RM_0$ (portfolio) & $223.664$ \\
\bottomrule
\end{tabular}
\end{table}

\begin{figure}[!htbp]
    \centering
    \includegraphics[width=0.80\textwidth]{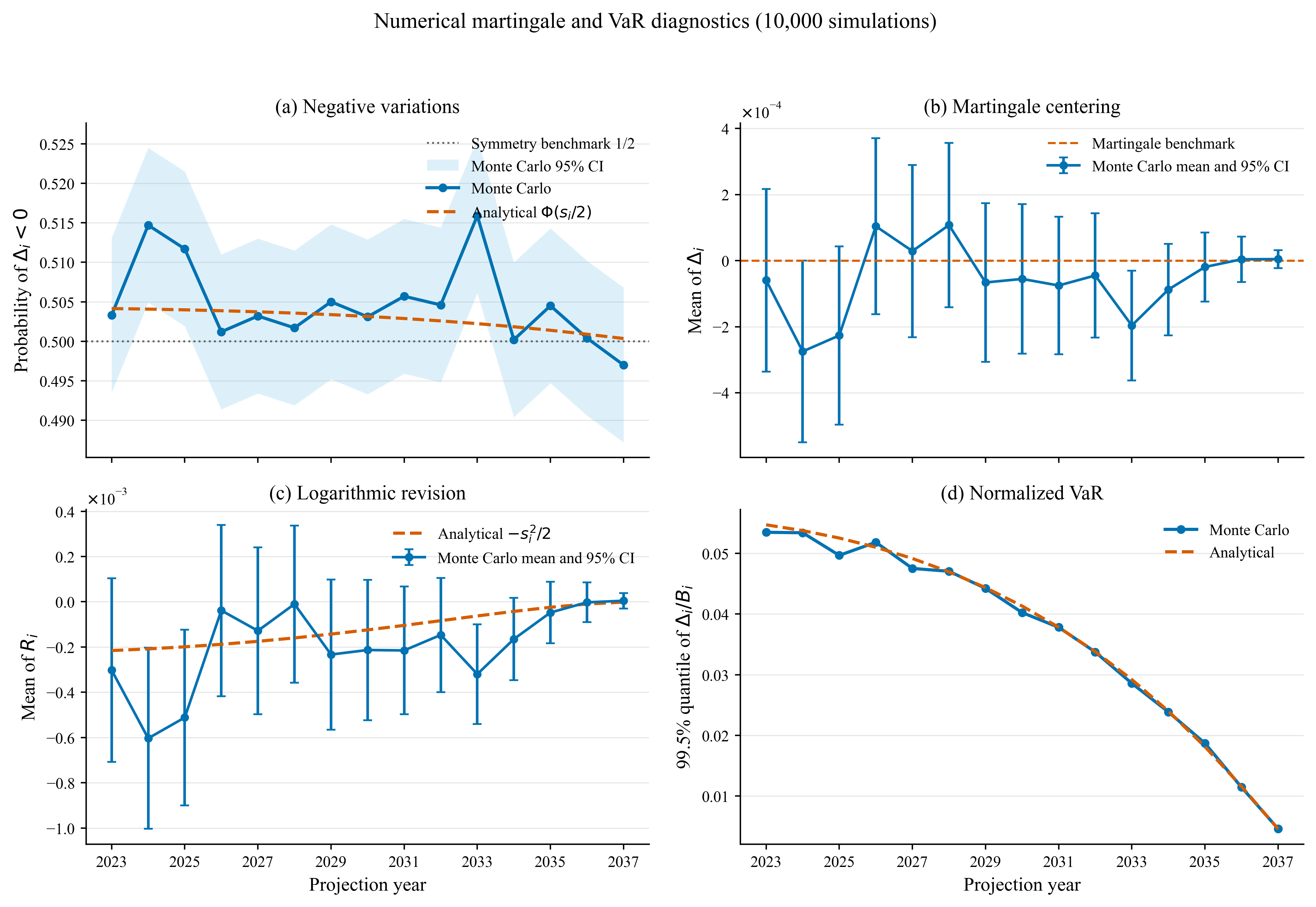}
    \caption{Numerical consistency checks of the conditional one-year Best Estimate revision distribution: (a) frequency of negative variations, (b) mean discounted variation with 95\% confidence intervals, (c) mean logarithmic revision with 95\% confidence intervals, and (d) analytical and simulated 99.5\% quantiles of the normalized variation.}
    \label{fig:analytical_validation}
\end{figure}

The distributional properties established in
Proposition~\ref{prop:conditional_be_var} are examined using 10,000
Monte Carlo simulations. Figure~\ref{fig:analytical_validation} compares
simulated frequencies, moments and normalized quantiles with their
analytical counterparts. These are numerical consistency checks of the
conditional distribution underlying the analytical SCR, rather than a
direct validation of the aggregated Risk Margin.

The empirical frequencies of negative valuation changes remain close to the theoretical probabilities $\Phi(s_i/2)$, with a maximum absolute gap of approximately $0.0137$. Since the implied deviation from one half is small relative to Monte Carlo sampling uncertainty, individual empirical frequencies may fall below one half without contradicting the theoretical asymmetry.

The 95\% confidence intervals for the simulated means of $\Delta_i$ contain zero at 14 of the 15 non-degenerate projection dates. The single exception is small in magnitude and is consistent with finite-sample Monte Carlo variation. The mean logarithmic revisions are likewise close to their theoretical values $-s_i^2/2$. Moreover, the analytical and simulated 99.5\% quantiles of the normalized variations remain close throughout the projection horizon, with a mean absolute relative gap of approximately $1.61\%$. In particular, the normalized quantile comparison checks the numerical implementation of the conditional VaR formula; it does not independently establish the aggregate analytical Risk Margin.

\medskip

The sensitivity of the portfolio Risk Margin to age and
maturity is examined in Table~\ref{tab:sensitivity_rm}.
The analytical Risk Margins are taken as reference values
and compared with conditional Monte Carlo estimates
of the same projected capital requirements.

\begin{table}[htbp]
    \centering
    \caption{Analytical and conditional Monte Carlo
    VaR-based Risk Margins across age--maturity
    configurations, using a 6\% Cost-of-Capital rate
    (portfolios of 10,000 contracts).}
    \label{tab:sensitivity_rm}
    \renewcommand{\arraystretch}{1.12}
    \begin{tabular}{ccc}
        \toprule
        \textbf{Age / Maturity}
        & \textbf{Analytical}
        & \textbf{Conditional MC} \\
        \midrule
        $60/15$ & $223.664$ & $223.535$ \\
        $65/10$ & $130.530$ & $130.482$ \\
        $75/10$ & $192.220$ & $192.079$ \\
        $80/10$ & $193.166$ & $192.956$ \\
        \bottomrule
    \end{tabular}
\end{table}
\FloatBarrier

The conditional Monte Carlo estimates closely agree
with their analytical counterparts, with relative
discrepancies ranging from $-0.036\%$ to $-0.109\%$.
This agreement provides a numerical consistency
check of the projected conditional capital calculations.\\

For a fixed maturity of 10 years, the analytical
Risk Margin increases from $130.530$ at age 65
to $192.220$ at age 75 and $193.166$ at age 80.
The reference portfolio, aged 60 with a longer
maturity of 15 years, produces a Risk Margin
of $223.664$. Since both age and maturity differ,
this comparison does not isolate either effect.\\

Overall, the results highlight the joint influence
of age-specific mortality dynamics and the remaining
exposure horizon on longevity capital costs.
The analytical VaR-based Risk Margin provides
the reference measure for comparison with
the stress-based results in
Subsection~\ref{subsec:stress_pricing_results}.

\subsection{Stress-based pricing results}
\label{subsec:stress_pricing_results}

The two prospective stress specifications of Section~\ref{sec:structural_stress}
are evaluated using the same-date capital definition in
\eqref{eq:stress_projected_scr}. Table~\ref{tab:stress_results} reports the
resulting Risk Margins for exploratory shocks of $K=10\%,20\%,30\%$,
using a $6\%$ Cost-of-Capital rate.

\begin{table}[!htbp]
  \centering
  \caption{Scenario-based Risk Margins at a $6\%$ Cost-of-Capital rate
  (portfolio of 10,000 contracts).}
  \label{tab:stress_results}
  \renewcommand{\arraystretch}{1.17}
  \begin{tabular}{lccc}
    \toprule
    \textbf{Stress scenario} & \textbf{$K=10\%$} & \textbf{$K=20\%$} & \textbf{$K=30\%$} \\
    \midrule
    Mortality-level stress & $94.764$ & $191.361$ & $289.831$ \\
    Trend-based stress & $26.3224$ & $52.8230$ & $79.5034$ \\
    \bottomrule
  \end{tabular}
\end{table}
\FloatBarrier

The Risk Margin increases with shock severity in both cases. Mortality-level shocks produce larger values because they directly affect prospective mortality intensities, whereas trend-based shocks operate through the latent longevity factor.

Under the estimated parameters, the deterministic target
$\beta(t)$ has effectively converged to its long-run level $\bar{\theta}$ by the 2023 valuation date. Consequently, the trend-based stress primarily represents a persistent downward revision of this long-run target, rather than a change in the convergence speed $\gamma$. Its effect on future mortality intensities is transmitted through the model dynamics.

\begin{figure}[!htbp]
  \centering
  \includegraphics[width=0.94\textwidth]{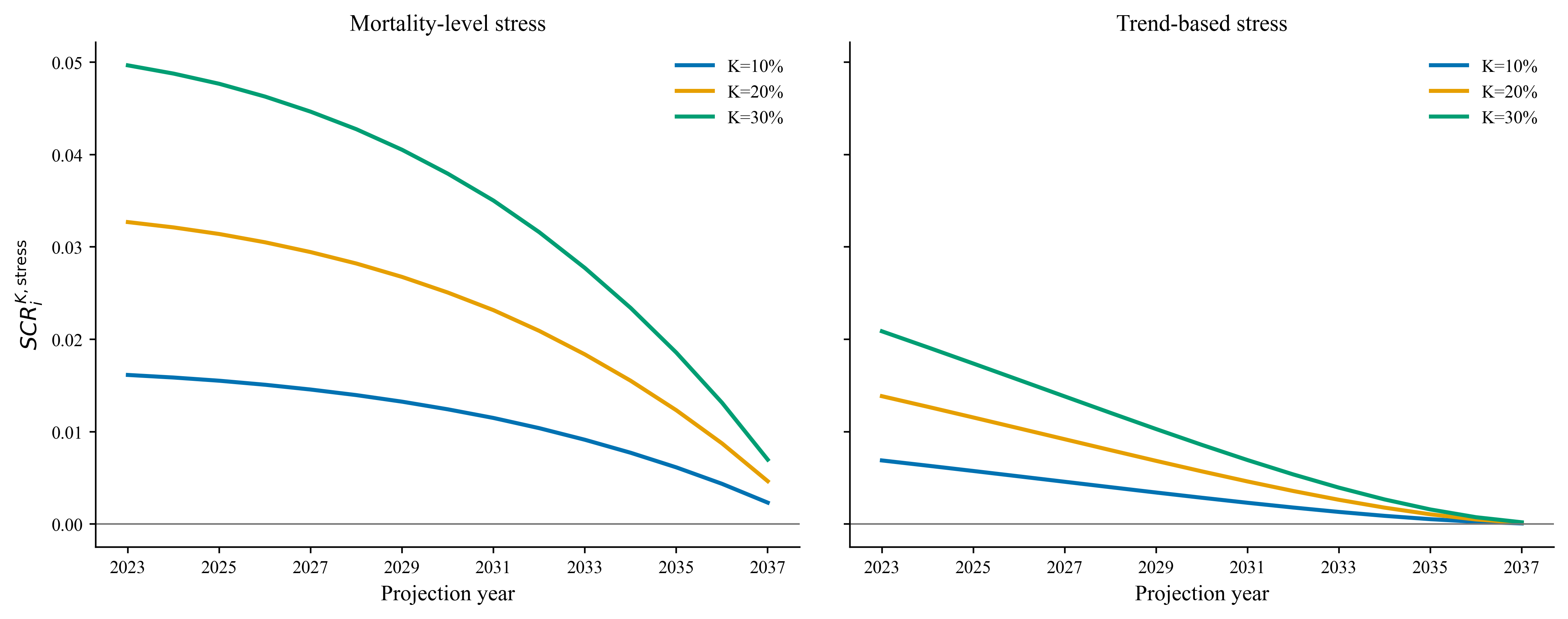}
  \caption{Projected scenario-based capital requirements per contract
  under mortality-level and trend-based stresses for $K=10\%,20\%,30\%$
  (2023--2037).}
  \label{fig:stress_scr_profiles}
\end{figure}

Figure~\ref{fig:stress_scr_profiles} shows the projected capital
requirements per contract before Cost-of-Capital weighting. The capital
profiles decrease towards maturity and increase with shock intensity,
with the mortality-level stress producing the larger initial exposure.

The discounted aggregation of these capital profiles gives the Risk Margins
in Table~\ref{tab:stress_results}.

\FloatBarrier

\subsection{Numerical stability and actuarial interpretation}
\label{subsec:numerical_stability_interpretation}

Beyond the reported VaR-based capital requirements
and scenario-based capital measures, the numerical
exercise provides further evidence on numerical
robustness and the actuarial interpretation of
the results.\\

A first numerical point concerns the stability of the conditional Monte
Carlo estimate of the projected VaR-based capital requirements. For the
reference 60/15 portfolio, Appendix~\ref{app:rm_stability} reports values
from $223.086$ to $223.579$ for 500 to 10,000 simulations, with an estimate
of $223.535$ at 10,000 simulations compared with the analytical reference
of $223.664$. The analysis illustrates numerical stability of the
conditional Monte Carlo implementation; it is not a convergence test of
the separate ex-ante unconditional quantile measure.\\

The simulated cumulative survival trajectories reported in Appendix~\ref{app:simulated_survival_trajectories} provide a complementary view of the stochastic uncertainty generated by the affine mortality model. They illustrate how small annual variations in mortality intensities may accumulate over the projection horizon and affect the valuation of longevity-linked exposures. This cumulative effect is particularly relevant for annuity-related liabilities, whose value depends on the persistence of survival probabilities over time rather than on a single-period mortality shock.\\

An additional numerical diagnostic concerns the
non-negativity of mortality intensities under the
Gaussian affine specification. Since the model does
not impose positivity constraints, negative mortality
intensities may occur along simulated trajectories.

For the reference portfolio aged 60 with a 15-year
horizon, a diagnostic based on 10,000 simulations
shows that 0.335\% of the simulated mortality
intensities are negative across the considered
age-horizon grid. Only three negative annual
integrated mortality values are observed across
150,000 simulated intervals (0.002\%), while no
cumulative or analytical conditional survival
probabilities exceeding one are recorded.

To assess the impact of these violations on the
valuation, two numerical sensitivity experiments
are conducted. The first corrects negative annual
integrated mortality values, while the second
replaces negative intensities by zero when
evaluating survival quantities, retaining the
original Gaussian capital factors.

Relative to the same-path Monte Carlo reference,
the resulting changes in the aggregated capital-cost
estimate are $-0.0000035\%$ and $-0.0003519\%$,
respectively. These results indicate negligible sensitivity of the aggregated capital-cost estimate to the considered numerical adjustments, supporting the stability of the valuation for the reference portfolio.\\

From an actuarial perspective, the empirical results highlight three main points. First, the affine mortality model offers a tractable basis for Solvency II-based valuation because it combines stochastic mortality dynamics with closed-form conditional survival probabilities. This property is essential for repeated revaluation, since Best Estimate and SCR calculations must be performed at several future projection dates.\\

Second, the VaR-based Risk Margin and the scenario-based Risk Margins provide complementary views of structural longevity risk. The VaR-based assessment uses a high quantile of one-year Best Estimate revaluations, whereas scenario-based capital is defined from the non-negative expected same-date liability increase under explicitly specified adverse mortality assumptions. Both measures are aggregated using the Cost-of-Capital principle, but the scenario-based capital requirements are not regulatory SCRs and are not calibrated to the same probability level as the VaR measure.\\

Third, the trend-based stress provides an actuarial interpretation of structural longevity risk. Rather than applying an immediate proportional reduction to mortality intensities, it modifies the future path of mortality improvements through the parameters of the longevity trend. This formulation is particularly relevant for long-term annuity exposures, where persistent changes in survival
expectations affect liability valuation. The numerical results show that the trend-based stress produces substantially lower scenario-based Risk Margins than the mortality-level stress, reflecting the different ways in which the two specifications affect future mortality.\\

These findings provide numerical support for the analytical conditional VaR-based capital calculation and the actuarial interpretation of the stress-based results. They also show that the stress-based extension can be implemented without losing the tractability of the affine mortality setting, while clarifying how adverse longevity assumptions affect scenario-based capital valuation.

\section*{Conclusion}
\label{sec:conclusion}

This paper develops a Solvency II-based Cost-of-Capital framework for the prudential valuation of longevity-linked exposures in incomplete longevity markets. The approach combines a continuous-time affine mortality model with a Cost-of-Capital valuation framework, thereby providing a tractable link between stochastic mortality dynamics, survival-index payoffs, one-year capital requirements and prudential valuation.
 The use of an affine calendar-year mortality model is particularly useful in this setting, because it preserves closed-form conditional survival probabilities and makes repeated revaluation feasible along the projection horizon. The Gaussian affine structure also allows us to characterize the conditional distribution of one-year discounted Best Estimate variations. In particular, we establish their mean-zero property, the asymmetry of positive and negative variations, and an analytical expression for the corresponding Value-at-Risk. The resulting analytical Risk Margins are closely reproduced by conditional Monte Carlo estimates across the four age--maturity configurations, while the numerical diagnostics support the implementation of the underlying conditional distribution. \\

The analytical valuation results are complemented by a scenario-based assessment of structural longevity risk through prospective shocks to the deterministic reversion target of the latent longevity factor. Unlike a mortality-level stress, which directly rescales future intensities, this specification revises the future mean-reversion target while preserving the central state at each valuation date. The comparison illustrates how distinct stress transmission channels affect scenario-based capital-cost measures without assigning statistical probabilities to the selected shock magnitudes.\\

The empirical application to Belgian mortality data illustrates how the valuation framework operates in practice. 
The results show that the mortality-level stress
produces substantially higher scenario-based Risk Margins
than the trend-based stress, with both increasing
as stress intensity rises. This difference reflects the immediate impact
of the mortality-level stress, compared with the indirect transmission of changes in the latent longevity factor's long-run target under the trend-based specification. The results therefore highlight
the importance of the stress specification in the assessment of longevity risk and support the use of trend-based stress testing as a complement to conventional mortality-level shocks.\\

Several extensions may strengthen the framework further. First, the sensitivity of the deterministic longevity trend to the calibration window suggests that trend specification remains a relevant source of model risk. Future work could consider stochastic trend dynamics, regime-switching specifications or alternative filtering procedures to capture persistent changes in mortality improvements more flexibly. Second, the analysis could be extended to multi-population settings, where basis risk between a reference population and an insured portfolio becomes central for longevity risk transfer. Third, incorporating additional financial and demographic sources of uncertainty would allow a broader assessment of longevity-linked transactions under incomplete market conditions.\\

The resulting approach contributes to the actuarial valuation of longevity-linked exposures by linking affine mortality modelling, Cost-of-Capital pricing and explicitly specified scenario-based capital measures within a single prudential framework. These alternative stress-based Risk Margin measures complement, but should not be confused with, a regulatory Solvency II Risk Margin. 
It highlights that the prudential valuation of
longevity-linked exposures depends not only on
the magnitude of longevity risk, but also on how
adverse changes in mortality improvements affect
future survival expectations. This perspective provides a useful basis for future work on the pricing, transfer and regulatory assessment of long-term longevity risk.


\appendix

\section{Additional numerical and methodological details}
\label{app:additional_details}

This appendix reports complementary numerical elements supporting the empirical application developed in Section~\ref{sec:empirical_application}. The material is not intended to introduce additional methodological assumptions. It provides supporting details on the volatility calibration, the computation of closed-form survival probabilities, simulated survival trajectories and the numerical stability of the portfolio Risk Margin.\\

\subsection{Calibration of the volatility function}
\label{app:volatility_calibration}

\begin{figure}[htbp]
	\centering
	\includegraphics[width=0.72\linewidth]{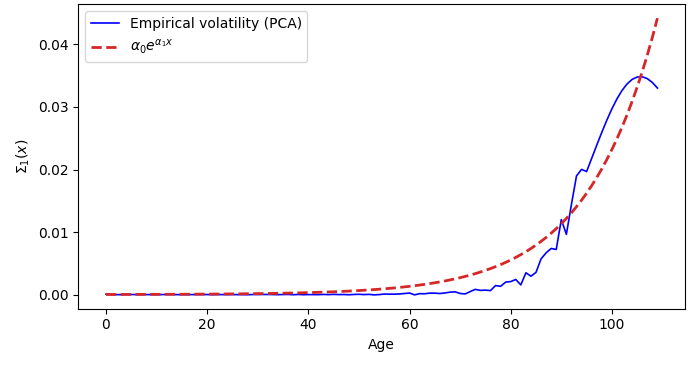}
	\caption{Calibration of the volatility function.}
	\label{fig:append_vol_calibration}
\end{figure}

Figure~\ref{fig:append_vol_calibration} compares the empirical volatility profile obtained from the principal component analysis with the fitted exponential specification used in the affine mortality model.\\

The fitted curve confirms the increasing age profile of mortality volatility. The empirical profile remains low over most ages and increases more sharply at advanced ages. This supports the use of an exponential volatility specification, which captures the main age-dependent pattern while preserving the analytical tractability required for closed-form survival probabilities and repeated valuation calculations.\\

\subsection{Sensitivity of the deterministic trend to the calibration period}
\label{app:trend_sensitivity}
The sensitivity of the deterministic trend is examined by re-estimating $\theta_0$, $\bar{\theta}$ and $\gamma$ over five historical windows ending in 2022. To isolate the effect of the calibration period on the deterministic component, all remaining model parameters are held fixed at their 1970--2022 reference estimates.\\

\begin{table}[htbp]
    \centering
    \caption{Sensitivity of the calibrated deterministic trend
    parameters to the calibration period.}
    \label{tab:trend_calibration_sensitivity}

    \renewcommand{\arraystretch}{1.15}
    \begin{tabular}{lccc}
        \toprule
        \textbf{Calibration period}
        & $\boldsymbol{\theta_0}$
        & $\boldsymbol{\bar{\theta}}$
        & $\boldsymbol{\gamma}$ \\
        \midrule
       1970--2022 & 0.414880 & 0.193305& 
       0.945240 \\
        1980-2022 & 0.193714 & 0.193602 & 0.999987\\
        1990-2022 & 0.202714 & 0.196233 & 0.230606 \\
        2000-2022 & 0.271639 & 0.190951 & 0.608378 \\
        2010-2022 & 0.205759 & 0.205754 & 0.194752 \\
   \bottomrule
    \end{tabular}   
\end{table}
\FloatBarrier

\begin{figure}[!htbp]
\centering
\includegraphics[width=0.70\textwidth]{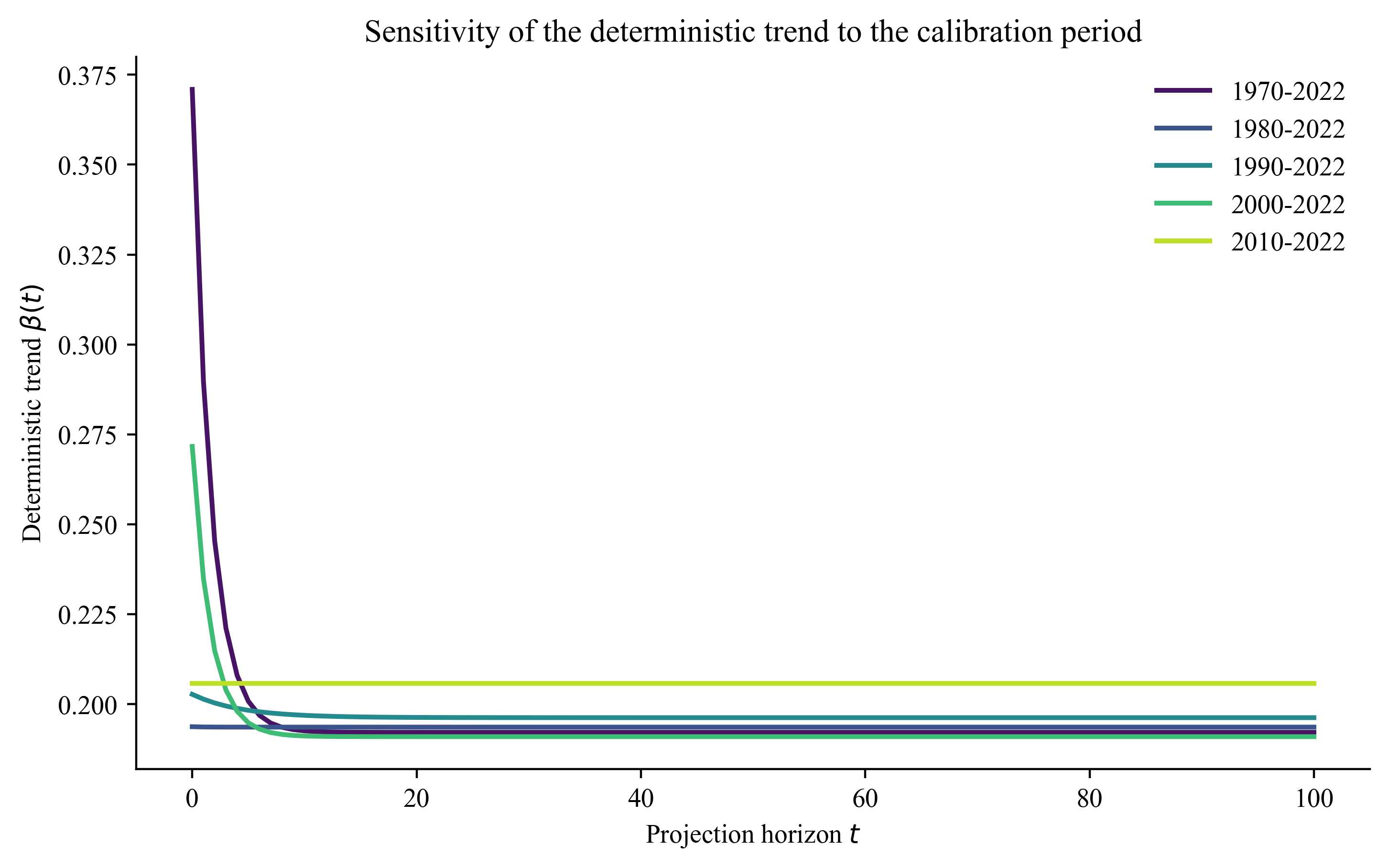}
\caption{Sensitivity of the deterministic trend function $\beta(t)$ to the calibration period.}
\label{fig:trend_calibration_sensitivity}
\end{figure}

Under the imposed parameter constraints, all estimated deterministic trends are monotonically decreasing.
As shown in Figure~\ref{fig:trend_calibration_sensitivity}, the
calibration period mainly affects the initial level and the rate of adjustment of the deterministic trend. Despite differences in their short-term trajectories, the projected trends approach relatively close long-run levels, ranging from approximately
0.191 to 0.206.

The sensitivity to the calibration window is therefore most pronounced over the early projection years, rather than over long horizons. Some trajectories are nearly flat, reflecting a limited difference between the estimated initial and long-run levels or a rapid convergence towards the latter.

\subsection{Illustration of closed-form survival probabilities}
\label{app:closed_form_survival}

Table~\ref{tab:survival_probabilities} reports the
Monte Carlo averages of scenario-conditional survival
probabilities under the proposed stochastic mortality
model for an individual aged 60 in 2023.
 The remaining maturity horizon is denoted by $\xi$, while ${}_{\xi}m_{t,x}$ and ${}_{\xi}v_{t,x}$
correspond respectively to the mean and variance terms entering the closed-form survival probability formula.

\begin{table}[!htbp]
    \centering
    \caption{Mean scenario-conditional survival probabilities
    under the stochastic mortality model.}
    \label{tab:survival_probabilities}
    \renewcommand{\arraystretch}{1.10}
    \small
    \begin{tabular}{ccccc}
        \toprule
        Year & $t_{\mathrm{model}}$ & Age & $\xi$
        & Mean survival probability \\
        \midrule
       2023 & 53 & 60 & 15 & 0.79113730 \\
        2024 & 54 & 61 & 14 & 0.79630295 \\
        2025 & 55 & 62 & 13 & 0.80191196 \\
        2026 & 56 & 63 & 12 & 0.80858236 \\
        2027 & 57 & 64 & 11 & 0.81602153 \\
        2028 & 58 & 65 & 10 & 0.82442274 \\
        2029 & 59 & 66 &  9 & 0.83416121 \\
        2030 & 60 & 67 &  8 & 0.84495431 \\
        2031 & 61 & 68 &  7 & 0.85689359 \\
        2032 & 62 & 69 &  6 & 0.87075495 \\
        2033 & 63 & 70 &  5 & 0.88636060 \\
        2034 & 64 & 71 &  4 & 0.90379167 \\
        2035 & 65 & 72 &  3 & 0.92332941 \\
        2036 & 66 & 73 &  2 & 0.94552491 \\
        2037 & 67 & 74 &  1 & 0.97090975 \\
        2038 & 68 & 75 &  0 & 1.00000000 \\
        \bottomrule
    \end{tabular}
\end{table}
\FloatBarrier

As the remaining maturity $\xi$ decreases, the mean
scenario-conditional survival probability increases
and reaches one at maturity. These probabilities are obtained by averaging scenario-conditional survival probabilities across Monte Carlo simulations. The underlying scenario-conditional probabilities are also used in the Best Estimate revaluations underlying
the SCR and Risk Margin calculations.

\subsection{Simulated cumulative survival trajectories}
\label{app:simulated_survival_trajectories}

\begin{figure}[!htbp]
    \centering
    \includegraphics[width=0.82\textwidth]
    {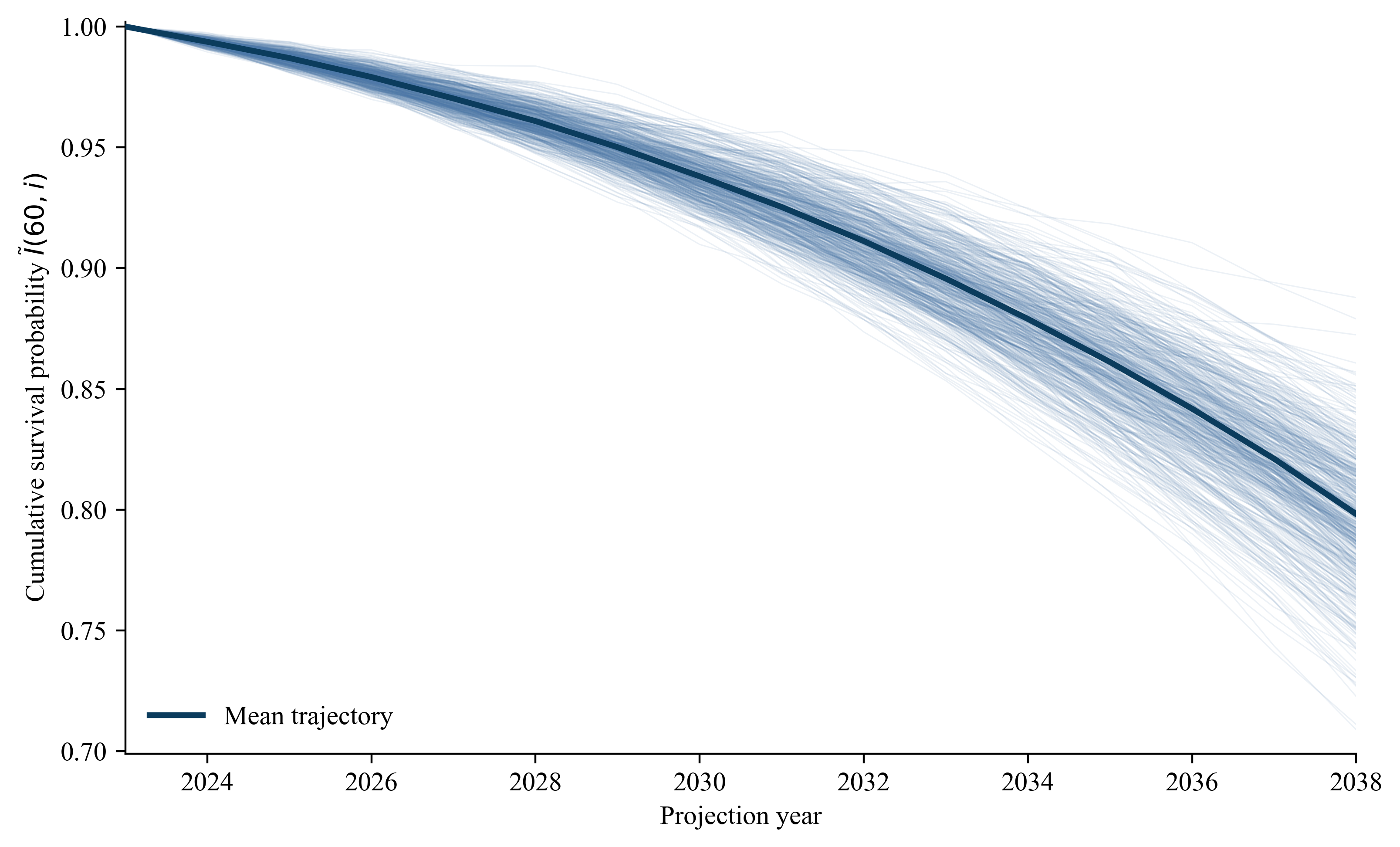}
    \caption{Simulated cumulative survival trajectories for the cohort aged 60 over a 15-year horizon. The thin curves represent 500 trajectories selected from 10,000 simulations, while the thick curve represents the mean trajectory.}
    \label{fig:simulated_survival_60_15}
\end{figure}

Figure~\ref{fig:simulated_survival_60_15} displays simulated cumulative survival trajectories generated under the stochastic affine mortality model.\\


The trajectories illustrate the dispersion generated by stochastic mortality dynamics over the projection horizon. This dispersion is the main source of uncertainty entering the simulated one-year liability variations used in the VaR-based capital assessment. From an actuarial perspective, the figure shows how small annual changes in mortality intensities may accumulate into non-negligible differences in survival values over time.\\

\subsection{Monte Carlo stability of the portfolio Risk Margin}
\label{app:rm_stability}

Table~\ref{tab:rm_stability} reports the stability of the conditional Monte Carlo Risk Margin for the reference portfolio $(x_0,T)=(60,15)$, alongside its analytical benchmark.

\begin{table}[!htbp]
	\centering
	\caption{Conditional Monte Carlo stability for the 60/15 portfolio (6\% Cost-of-Capital rate).}
	\label{tab:rm_stability}
	\begin{tabular}{cc}
		\toprule
		\textbf{Number of simulations} & \textbf{Portfolio Risk Margin} \\
		\midrule
		500    & 223.086 \\
		1,000  & 223.187 \\
		2,000  & 223.433 \\
		5,000  & 223.579 \\
		10,000 & 223.535 \\
		\midrule
		Analytical reference & 223.664 \\
		\bottomrule
	\end{tabular}
\end{table}

The conditional Monte Carlo estimate is $223.535$ at 10,000 simulations, compared with the analytical value of $223.664$ (a relative difference of approximately $-0.058\%$). The results are stable at the larger simulation sizes; they do not provide confidence intervals across independent simulation seeds.

\subsection{Sensitivity to the revised Solvency II Risk Margin framework}
\label{app:nv_reforme}

As a supplementary sensitivity analysis, we examine the
impact of the revised Solvency II framework applicable from 30 January 2027. Commission Delegated Regulation (EU) 2026/269 \cite{EUDelegated2026} introduces a lower Cost-of-Capital rate of 4.75\% and a time-dependent attenuation factor applied to projected capital requirements.

Using the same projected annual capital requirements as in the baseline analysis, the revised aggregation is given by
\begin{equation}
\label{eq:rm_revised_2027}
RM_0^{2027}
=
0.0475
\sum_{i=0}^{T-1}
P(0,i+1)
\,\max\left(0.96^i,\,0.50\right)
SCR_{i|0}.
\end{equation}

The attenuation factor progressively reduces the weight
assigned to future capital requirements, subject to a
minimum value of 50\%. All three columns below use the same projected expected conditional SCRs; only the Cost-of-Capital rate and the attenuation weights change. The intermediate 4.75\% column without attenuation isolates the rate effect.

Table~\ref{tab:rm_reform_sensitivity} reports the corresponding analytical conditional VaR-based Risk Margins for the four age--maturity configurations.

\begin{table}[htbp]
\centering
\caption{Sensitivity of analytical VaR-based Risk Margins to the revised Solvency II aggregation (portfolios of 10,000 contracts).}
\label{tab:rm_reform_sensitivity}
\renewcommand{\arraystretch}{1.12}
\begin{tabular}{cccc}
\toprule
\textbf{Age / Maturity} & \textbf{6\%, no attenuation} & \textbf{4.75\%, no attenuation} & \textbf{4.75\%, attenuated} \\
\midrule
60 / 15 & 223.664 & 177.067 & 144.630 \\
65 / 10 & 130.530 & 103.336 & 91.223 \\
75 / 10 & 192.220 & 152.174 & 134.232 \\
80 / 10 & 193.166 & 152.923 & 134.862 \\
\bottomrule
\end{tabular}
\end{table}

The 2027 aggregation reduces the analytical Risk Margin for the reference 60/15 portfolio from $223.664$ to $144.630$ (approximately $35.3\%$). The intermediate column distinguishes the impact of the lower Cost-of-Capital rate from that of the attenuation factor. The underlying mortality model and capital profiles remain unchanged.

The same comparison is performed for the alternative scenario-based capital requirements defined in \eqref{eq:stress_projected_scr}. For each scenario, the projected capital profile is kept fixed and only the Cost-of-Capital aggregation weights are modified. These figures are \emph{not} regulatory SCR results; they illustrate the sensitivity of the alternative stress-based Risk Margin measure to the revised aggregation convention.

\begin{table}[htbp]
\centering
\caption{Sensitivity of scenario-based Risk Margins to the
revised 2027 Cost-of-Capital framework for the reference
60/15 portfolio of 10,000 contracts.}
\label{tab:rm_stress_reform_sensitivity}

\renewcommand{\arraystretch}{1.18}
\begin{tabular}{llccc}
\toprule
\textbf{Stress scenario}
& \textbf{CoC framework}
& \multicolumn{3}{c}{\textbf{Stress intensity}} \\
\cmidrule(lr){3-5}
& & \textbf{$K=10\%$}
  & \textbf{$K=20\%$}
  & \textbf{$K=30\%$} \\
\midrule

\multirow{2}{*}{Mortality-level stress}
& Baseline (6\%)    & 94.764 & 191.361 & 289.831 \\
& Revised (2027)   & 61.220 & 123.672 & 187.383 \\

\addlinespace[5pt]

\multirow{2}{*}{Trend-based stress}
& Baseline (6\%)    & 26.3224 & 52.8230 & 79.5034 \\
& Revised (2027)   & 17.9881 & 36.1039 & 54.3485 \\

\bottomrule
\end{tabular}
\end{table}

Across both stress families, the revised weights lower the resulting cost-of-capital valuation. This comparison isolates the change in aggregation and does not imply regulatory equivalence of the underlying scenarios.

\FloatBarrier

\section*{Statements and Declarations}

\noindent\textbf{Funding.}
The authors received no specific funding for this research.

\medskip
\noindent\textbf{Competing interests.}
The authors declare that they have no competing
interests relevant to the content of this article.

\medskip
\noindent\textbf{Data availability.}
The Belgian mortality data used in this study
were obtained from the Human Mortality Database
(\url{https://www.mortality.org/}), subject to
its conditions of access and use.

\medskip
\noindent\textbf{Code availability.}
The Python code used for model calibration,
simulation, and numerical analysis is available
from the corresponding author upon reasonable
request.

\medskip
\noindent\textbf{Author contributions.}
Madiha Bouhouche contributed to the methodology, software implementation, numerical analysis, and preparation of the original manuscript.

\noindent Khalil Said contributed to the methodological development, scientific supervision, and critical review of the manuscript.

\noindent Fadoua Badaoui contributed to the scientific review and editing of the manuscript.

\noindent All authors reviewed and approved the final version of the manuscript.

\medskip
\noindent\textbf{Ethics approval and consent
to participate.}
Not applicable. This study uses aggregated
population-level mortality data and does not
involve direct participation of human subjects.

\medskip
\noindent\textbf{Consent for publication.}
Not applicable.

\section*{AI Use Declaration}

ChatGPT (OpenAI) was used to assist with language editing,
LaTeX formatting, and improving the clarity of selected
passages. The authors reviewed the manuscript and retain
full responsibility for its scientific content.

\bibliography{Bib-RL}

\end{document}